# Optical Deformability of Fluid Interfaces

**Jean-Pierre DELVILLE[*], Alexis CASNER[†], Régis WUNENBURGER**

Centre de Physique Moléculaire Optique et Hertzienne, UMR CNRS/Université No 5798, Université Bordeaux I, 351 Cours de la Libération, F-33405 Talence cedex, France.

**Iver BREVIK[#]**

Department of Energy and Process Engineering, Norwegian University of Science and Technology, N-7491 Trondheim, Norway.

**Abstract:**

The formation, deformation, and break-up of liquid interfaces are ubiquitous phenomena in nature. In the present article we discuss the deformation of a liquid interface produced by optical radiation forces. Usually, the bending of such an interface by the radiation pressure of a c.w. laser beam is weak. However, the effect can be enhanced significantly if one works with a near-critical phase-separated liquid mixture, whereby the surface tension becomes weak. The bending may in this way become as large as several tenths of micrometers, even with the use of only moderate laser power. This near-criticality is a key element in our experimental investigations as reviewed in the article. The effect is achieved by working with a micellar phase of microemulsions, at room temperature. We give a brief survey of the theory of electromagnetic forces on continuous matter, and survey earlier experiments in this area, such as the Ashkin-Dziedzic optical radiation force experiment on a water/air surface (1973), the Zhang-Chang experiment on the laser-induced deformation of a micrometer-sized spherical water droplet (1988), and the experiment of Sakai et al. measuring surface tensions of interfaces in a non-contact manner (2001). Thereafter, we survey results we obtained in recent years by performing experiments on near-critical interfaces, such as interface bending in the linear regime, stationary large deformations of liquid interfaces, asymmetric pressure effects on interfaces under intense illumination, nonlinear deformations, and laser-sustained liquid columns.

---

[*] Electronic address: jp.delville@cpmoh.u-bordeaux1.fr

[†] Present address: Département de Conception et de Réalisation d'Expériences, CEA/DAM Ile-de-France, BP 12, F-91680 Bruyères-le-Châtel, France. Electronic address: alexis.casner@cea.fr

[#] Electronic address: iver.h.brevik@ntnu.no

I - Introduction

Formation, deformation and break-up of fluid interfaces are ubiquitous phenomena in daily life that play a significant role in science and technology [1]. From the fundamental point of view they also illustrate a fascinating behavior called finite time singularity [2], where the most common and famous example is the break up of pendant drops driven by gravity [3, 4]. Extension to drop deformation under an applied field proved to be even richer due to the addition of an external control parameter. For example, interface instabilities driven by electric fields [5, 6] represent nowadays the corner stone of many industrial processes as different as electro-spraying [7], ink-jet printing [8] or surface relief patterning [9]. A uniform magnetic field can as well destabilize fluid interfaces to create well-organized peak structures [10, 11] or to form elongated magnetic droplets [12]. These deformations, as well as those induced by the acoustic radiation pressure on liquid surfaces [13, 14], were essentially used to explore and characterize in a non-contact way, the mechanical properties of fluid interfaces [15, 16].

Surprisingly, such a strategy has been extended to the optical deformation of soft materials only recently. Indeed, while the interaction between a laser wave and a micrometer-sized dielectric particle has attracted considerable interest in recent years, and most advances were devoted to optical levitation and trapping (for a recent review see for example [17]), much less attention appears to have been directed to the deformation of soft transparent interfaces despite its large field of practical applications (see below). The main reason is linked to the fact that deformation of fluid interfaces by the optical radiation is intrinsically weak. It is also very sensitive to secondary disturbing effects associated to the optical absorption of the fluids (laser heating, convection or thermocapillary effects or instance).

The present survey is devoted to the analysis of this coupling between interface and laser waves. We theoretically describe and experimentally explore laser-induced liquid interface deformations driven by radiation pressure. As optical radiation pressure effects are weak, experimental investigations were performed on the meniscus of phase separated liquid mixtures close to a liquid-liquid critical point. Indeed, since surface tension vanishes close to a critical point, near-critical interfaces become highly deformable. Moreover, as surface tension, and density and refractive index contrasts are continuous functions of the temperature shift to the critical temperature, it becomes possible to tune the fluid properties by working with a single medium and simply adjusting one parameter. Finally, close to a critical point most liquids belong to the universality class $(d = 3, n = 1)$ of the Ising model, where $d$ is the space dimension involved, and $n$ the dimension of the order parameter (here the density contrast). Our results can thus be generalized to any fluids belonging to the same universality class.

In these conditions, huge stationary interface deformations of several tens of microns can be induced with low beam c.w. lasers. The continuous variation of the surface softness by a temperature scanning also allows a universal description of the phenomenon because it becomes possible to overlap the range of variation of the excitation and the hydrodynamic length scales, respectively given by the beam waist and the capillary length. New predicted behaviors in linear interface deformation are then seen to occur, particularly the transition from a non-local to a local excitation of the interface when the critical point is neared. All the data can then be cast on a single master curve when rescaled with an optical Bond number. As interface bending by radiation pressure is expected to be independent on the direction of the beam propagation, we also generalize our universal description to both upward and downward exciting beams.

On the other hand, as surface tension can becomes very weak, we also investigated nonlinear regimes in deformation. We show that deformations become asymmetric versus beam propagation. While surprising stationary tether shapes are observed when the exciting beam crosses the interface from the low to the large refractive index liquid, the induced hump can become unstable for sufficiently large beam intensities in the opposite case. We analyze this new opto-hydrodynamic instability, by exploring the universal behavior of its onset. When properly rescaled, the dispersion in measured onsets, versus temperature shift and beam waist, also reduces to a single master behavior, which is retrieved theoretically from simple arguments. Above onset, the optically driven interface instability leads to the formation of stationary beam-centered liquid micro-jet emitting droplets, which anticipates the bases for new applications in microfluidics and liquid micro-spraying. Finally, when working in finite geometry, the induced liquid jets lead to the formation of stable liquid columns of very large aspect ratio that would be unstable otherwise. These laser-induced liquid columns behave as reconfigurable self-written optical waveguides and provide opportunities for building new microfluidic-based devices for applications in adaptative optics.

The review is organized as follows. Before discussing our results, we present in Section II a well-documented survey on optical manipulation of liquid interfaces. As a theoretical description of laser induced interface deformation by radiation pressure is of great importance to analyze experimental results, we give a general description of electromagnetic forces on dielectrics in Section III, and show how a laser wave can couple to a liquid interface and drive its bending. Section IV is an experimental section. The fluid media used as well as the experimental setup is presented. Our investigation of interface bending by the radiation pressure starts in Section V. We investigate the linear regime in deformation for both upward and downward beams, we show that associated deformations are symmetric, and give a universal description of the involved phenomenon. We finally use interface deformation as adaptative lenses actuated by laser light. Our investigation of nonlinear behaviors begins in Section VI. We show that tether-like stationary deformations, instead of continuous elongation, are generated when the beam intercepts the interface from the fluid of lower refractive index. Consequently, the symmetry observed for linear deformations is broken at large beam intensities, since the beam can either induce a tether-like deformation or a liquid jet depending on the direction of propagation. This "opto-hydrodynamic" instability is analyzed, particularly the universal behavior of jetting onset. A comparison with already published data is also performed. Section VII is devoted to an extension of the previous one. As experiments are always performed in finite size volume, we show how liquid jetting driven by radiation pressure can be used to form and stabilize liquid columns of very large aspect ratio. Since these liquid columns have a larger refractive index than the surrounding fluid, they behave as reconfigurable adpatative optical fibers. This optical application as well as further potentialities in microfluidics is presented in Section VIII. We finally conclude in Section IX. Considering the good agreement observed between theory and experiments, as well as the diversity in produced liquid structures, the present review "rehabilitates" optical surface forces for further appealing investigations and applications that require micrometric length scale contact less manipulation of liquid interfaces.

## II - Optical Manipulation of Fluid Interfaces: a Brief Overview

### II.1 – Indirect Laser-Induced Interface Deformation

Since their invention in 1986, optical tweezers [18] have been recognized as a fundamental tool to manipulate micro-objects. An obvious extension going one-step forward optical trapping itself is to use the trapped particle either as a probe of the surrounding environment or as an actuator to constrain this environment. The first aspect is often related to the so-called optical dynamometry [19], while the second type of application has been widely used to induce an optical non-contact forcing on the studied structure in order to analyze how it adapts to this forcing or how it relaxes to its initial state after removing optical excitation. The study of red blood cells (RBC) is particularly illustrative because RBC's are highly deformable. For example Block and co-workers [20] tweeze a RBC against flow in order to analyze the resulting cell conformation. Cell relaxation from a parachute shape was measured on RBC deformed by a triple optical tweezers [21]. A third experiment was performed by attaching two beads to the RBC membrane, to trap optically these beads and to stretch indirectly the cell by pulling one beam from the other [22]. While this experiment gives access to an order of magnitude of the shear modulus of RBC's, it is nevertheless limited due to the fact that the bead sticking locally modifies the membrane properties. Another method consists in encapsulating a bead in a biologic membrane to form a tether by pulling it from the cell center [23]. Such an experiment gives access to the membrane viscosity. Conversely, one can exploit the relaxation of this tether. For example Steffen and co-workers [24] formed a long filament by stretching the interface between two coexisting phases in a Langmuir monolayer. They deduced the value of the line tension between these two phases by analyzing the tether relaxation. All these experiments clearly show how the use of an intermediate bead by optical tweezing allows the deformation of soft objects and the characterization of their surface property. However, since the trapped beads also bring difficulties in the interpretation of the results, a step forward in interface deformation analysis has been to remove this intermediate element in order to try to excite directly a membrane with a laser beam. While a major limitation in using directly focused beams on biologic objects is the excess overheating resulting from light absorption, a point which explains the use of intermediate beads, lasers can also alter membranes by simple optical interaction. Indeed, when a beam is highly focused on a deformable membrane, mechanical effects of light tend to attract large refractive index matter toward the focus. This effect locally induces excess in surface tension of the membrane that is responsible to the so-called "pearling instability" of cylindrical vesicles [25]. Local bilayer separation within the membrane or micro-vesicle expulsion by laser-induced tension within the mother's membrane can also be generated. All these phenomena, and particularly their dynamical aspect, were described in a review on laser-membrane interaction [26] and were theoretically interpreted [27]. Consequently, an intermediate bead is not always necessary to deform a soft interface. Direct effects of light beams on membranes are also observable and measurable, even in strongly nonlinear regimes. We will follow this direction in order to show how direct radiation pressure effects can be fully exploited.

### *II.2 – Direct Laser-Induced Interface Deformation: Optical Aspects*

While the first experiment on deformation of liquid interface by the radiation pressure of a laser wave [28] is contemporary to the beginning of experiments on optical levitation (they were developed by the same authors), this optical manifestation did not received the same enthusiasm neither development, until very recently. To explain this unbalanced situation, we will underline the intrinsic difficulties encountered in performing this type of experiment as well as the different possibilities that can be used to overcome them.

#### *A Qualitative Explanation*

Considering some theoretical developments [29], the first motivation of Ashkin and Dziedzic was to determine in which direction occurs the bending of a fluid interface separating two dielectric liquids of different refractive indices. According to other researchers this experiment could also bring some new insights on the right formulation of the electromagnetic energy-momentum tensor in dielectrics, either the Minkowski or the Abraham expression (see Section III). Even if the direction of laser-induced interface deformation cannot answer this question, the result obtained by Ashkin and Dziedzic was sufficiently surprising and non intuitive to present a qualitative interpretation.

Let us consider a laser beam at normal incidence to the fluid interface between two dielectric liquids 1 and 2, respectively characterized by the refractive index $n_1$ and $n_2$ (Figure II.1). We assume an ascending beam propagating along the $\hat{z}$ direction and $n_1 < n_2$ to be in agreement with the experiments presented in Section IV but of course, this particular case is just chosen to illustrate the phenomenon without removing any generality of our purpose. We can write momentum conservation for the system (photons+ interface) for an elementary surface element $dS$ during the time $dt$. If $\nu$ is the optical frequency, photon momentum $p$ in a medium of index of refraction $n$ can classically be written as $p = nh\nu/c$ [30]. Then momentum conservation leads to:

$$\frac{n_1 h\nu}{c} \hat{z} NSdt = \left( T \frac{n_2 h\nu}{c} - R \frac{n_1 h\nu}{c} \right) \hat{z} NSdt + d\vec{Q}_{int}, \tag{II.1}$$

where $N$ denotes the number of incident photons per surface and time unit, and $R$ and $T$ the reflection and transmission Fresnel coefficient in energy. By replacing $R$ and $T$ by their expressions $R = (n_1 - n_2/n_1 + n_2)^2$ and $T = 4n_1 n_2/(n_1 + n_2)^2$ at normal incidence, we finally find that momentum transferred to the interface per surface and time unit is:

$$\frac{1}{S} \frac{d\vec{Q}_{int}}{dt} = \frac{2n_1}{c} \left( \frac{n_1 - n_2}{n_1 + n_2} \right) Nh\nu \hat{z}. \tag{II.2}$$

Since the beam intensity $I$ is given by $I = Nh\nu$, the radiation pressure $\Pi_{Rad}$ exerted by the laser wave under the interface can be written as:

$$\Pi_{Rad} = \frac{2n_1}{c} \left( \frac{n_1 - n_2}{n_1 + n_2} \right) I \tag{II.3}$$

Eq. (3) shows that the associated force is generated by the photon refraction at the interface. However, the most surprising prediction is that interface bending does not necessarily occur in the direction of beam propagation. As we supposed $n_1 < n_2$, according to Eq. (2) the deformation is always directed towards the less refractive liquid, i.e. liquid 1 in our case, whatever the beam propagation is. Indeed, for a descending laser beam we should substitute $n_1$ by $n_2$ and $\hat{z}$ by $-\hat{z}$ which let Eq. (2) invariant. This means that interface bending does not depend on beam propagation, and the main reason for that is that photon gains momentum when passing from a low to a large refractive index medium. Ashkin and Dziedzic verified this remarkable property, which will be quantitatively illustrated in Section IV.

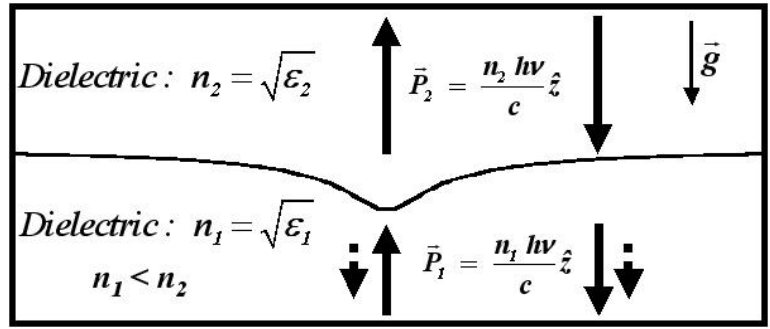

**Figure II.1:** Schematic of deformation of fluid interfaces by the optical radiation pressure. The left (resp. right) set of arrows illustrates momentum conservation for an ascending (resp. descending) laser beam; the dotted arrows represent momentum transferred to the interface.

*The experiment of Ashkin and Dziedzic*

Ashkin and Dziedzic [28] performed an experiment using a frequency doubled Nd:Yag laser (wavelength in vacuum $\lambda_0 = 0.53 \, \mu m$) strongly focused on the water/air interface (beam waist value $\omega_0 = 2.1 \, \mu m$). Due to the large value of the water/air surface tension ($\sigma = 73 \, mJ.m^{-2}$) they worked with laser pulses (pulse duration *60 ns* and peak power between 1 and *4 kW*) to increase radiation pressure effects. Since the interface bending is still very weak (typically a few tens of nanometers in height) with these excitation conditions, it is not directly observable. However, as an interface deformation behaves as a lens, one can deduce from the associated focal distance both the direction of the deformation and its height [29]. Ashkin and Dziedzic use these properties to demonstrate that interface deformation is always directed towards the medium of smaller index of refraction whatever the direction of beam propagation is. A quantitative interpretation of these experimental results is nevertheless difficult due to the spatial and temporal profiles of the laser pulses used. For example, in his theoretical interpretation, Brevik [31] had to consider a larger beam waist and to use a time delay to retrieve the temporal behavior of the induced lens. On the other hand, we should stress that very few results related to the deformation of fluid interfaces by the optical radiation pressure have been published. From the theoretical point of view, one can cite a description of Ashkin and Dziedzic experiment [32] and articles on the associated beam self-focusing in linear media [29, 33]. On the other hand, except a brief letter [34], we are aware of a second set of experiments realized at the liquid/air interface [35] using a ruby laser. As interface bending was still very weak, detection was performed by holographic interferometry. Nice agreement between theory and experiments is found but, as in Ashkin and Dziedzic experiment, deformations are not stationary and their weakness prevents any direct observation.

*Droplet Deformation and Optical microcavity*

Using a high speed camera, Zhang and Chang [36] performed the first direct observations of laser-induced interface deformation on spherical water droplets. The droplets (radius $a = 50 \ \mu m$) were illuminated by a dye laser (wavelength in vacuum $\lambda_0 = 0.6 \ \mu m$, pulse duration $0.4 \ \mu s$, and beam waist value $\omega_0 = 200 \ \mu m$). A first series of images shows oscillations of the interface after the passage of a *100 mJ* pulse. This series clearly evidenced that interface deformation is directed towards the air on both side of the drop with a small asymmetry on the exit face coming from the increase in intensity due to the ball lens effect of the drop. Even more surprising, they show that *200 mJ* pulses lead to the formation of a long liquid filament on the exit face of the drop and droplet disruption. Beyond the spectacular observations, these experiments were motivated by a research on the modification of the optical properties of dielectric spheres. Indeed, it is well known that drops illuminated by laser waves behave as optical microcavities in which whispering gallery modes and morphology dependent resonance are induced [37]. These properties can be tuned or altered by disturbing the shape of the drop [38]. Consequently, laser-induced liquid droplet deformation can be exploited to increase the optical coupling with the incident beam [39]. Finally, a quantitative interpretation of the Zhang and Chang experiment was given by Poon and coworkers [40] and further developed by Brevik [41]. Numerical simulations are in good agreement with experiments, even in the nonlinear regime in deformation.

*II.3 – Application of Laser-Induced Interface Deformation to Soft Matter Physics*

While optical bending by laser waves was clearly evidenced, the preceding paragraph also showed that the effect is very weak on classical interfaces because radiation pressure, which is proportional to the index contrast between the two liquids in contact, has to compete with the stabilizing effect of surface tension. This is certainly the reason why so few works have been dedicated to radiation pressure. However, the eighties have seen the emergence of the so-called soft matter physics that brought a sort of new birth to laser radiation pressure by opening new horizons beyond the optical physics area.

*Characterization of Weak Surface Tensions and Surface Waves*

Recently, a Japanese team exploited the bending of an interface by a laser wave to measure interfacial tensions [42] in a non-contact manner. Indeed, for particular experimental conditions, it will be shown in the following that the height of the deformation is inversely proportional to the surface tension of the fluid interface. As deformations are weak for classical interfaces, they measured, as Ashkin and Dziedzic, the associated lensing effect. Interface bending is induced by a c.w. $Ar^+$ pump laser (wavelength in vacuum $\lambda_0 = 0.514\ \mu m$, maximum power $P = 0.5\ W$) and probed with a He-Ne Laser ($\lambda_0 = 0.514\ \mu m$ $\lambda_0 = 0.633\ \mu m$) by analyzing far field diffraction. Note that for a beam power $P = 0.3\ W$ and a beam waist $\omega_0 = 142\ \mu m$, they found for the deformation of the water free surface a height of $2\ nm$! After a calibration of their set-up with the water free surface, they extended the technique to interfaces modified by adding a mystiric acid monolayer. Their measurements compared well with results obtained from the classical Wilhelmy's method. They also perform dynamic characterizations of interfaces. By modulating the pump beam, they showed that radiation pressure can excite surface waves, and they measured their power spectrum. In a second work [43], they focused their attention on the difficult case of ultra low surface tension, for which most classical techniques fail. Finally, they showed the pertinence of radiation pressure to excite high-frequency capillary waves [44], and measured shear viscosities.

*Characterization of Biological Membranes*

The characterization of the mechanical properties of lipidic membranes is of fundamental interest in biophysics. Since optical dynamometry is limited to large vesicles, Wang 's group used radiation pressure to quantitatively measure curvature rigidity of small vesicles [45] and in vivo cells [46] at a nanometric length scale (see above). Compared to optical dynamometry, these experiments show that the main interests in using radiation pressure are the strong localization of excitation as well as its non-invasive character

On the other hand, Käs' group patented a new tool to probe elasticity of cells using the radiation pressure. The apparatus, called "Optical Stretcher", stretches cells trapped by two beams propagating in opposite directions (c.w. Ti-Sapphire lasers working at $\lambda_0 = 0.780\ \mu m$). The advantage of this dual trap is a self-centring of the trapped cell associated to a symmetric radiation pressure effect on both cell hemispheres, in opposition to the Zhang and Chang experiment [36]. Consequently, the cell stretches along the beam axis as a soft dielectric medium [47]. The magnitude of the forces applied to the cell varies from the pico- to the nano-newton, i.e. typically one hundred times larger than that obtained by optical tweezing, by reducing radiation damage due to the fact that beams are not strongly focalized. Advantage of radiation pressure effect over stretching via attached beads trapped in a dual tweezers, is once more clearly evidenced. Due to its sensitivity, the ultimate goal of this technique is to be able to differentiate healthy to malignancy cells from the difference in elastic response [48]. Finally we should note that cells deform nonlinearly at large power excitation as illuminated droplets.

### II.4 – Summary

We showed in this section that direct radiation pressure effects are, in many situations, complementary to optical tweezing, particularly in soft matter physics where they are very efficient in characterizing fluid interfaces and membranes. However, although one of the most important advantages of the technique is that it is non-invasive, we have seen that little work has been devoted to the induced deformations. The reason is simple; discontinuity in photon momentum at an interface is weak and then, radiation pressure is also a weak effect that is not always observable. On the other hand, the use of high intensity lasers does not represent a significant advance in this area because an increase in intensity inevitably leads to strong thermal effect that usually dominates radiation pressure [49]. Consequently, radiation pressure deserves further investigations, and we will show in the following that appropriate interfaces allow for enhanced effects that are particularly appealing for the analysis of the general properties of radiation pressure effects: independence of deformation versus beam direction, associated lens effect, and non-linear deformations at high laser intensities. However, before discussing these manifestations, we need to clarify the theory devoted to mechanical effects of light in dielectric matter. This is the subject of the following section.

**III – Electromagnetic Forces on Dielectrics**

*III.1 – Basics*

We begin by writing down the commonly accepted expression for the electromagnetic volume force density $\vec{f}$ in an isotropic, nonconducting and nonmagnetic medium (see, for instance, [31, 41 50, 51]:

$$\vec{f} = -\frac{1}{2}|E|^2 \vec{\nabla}\varepsilon + \frac{1}{2}\vec{\nabla}\left[|E|^2 \rho \left(\frac{\partial \varepsilon}{\partial \rho}\right)_S\right] + \frac{\varsigma - 1}{c^2}\frac{\partial}{\partial t}\left(\vec{E} \times \vec{H}\right) \qquad \text{(III.1)}$$

Here $\vec{E}$ and $\vec{H}$ are the electric and magnetic fields, $\rho$ is the mass density of the medium (fluid), $\varepsilon$ is the permittivity, and $\varsigma = \varepsilon/\varepsilon_0$ is the relative permittivity where $\varepsilon_0$ denotes the permittivity of vacuum.

Let us comment on the various terms in Eq.(III.1), beginning with the last term. This term is called the Abraham term, since it follows from Abraham's electromagnetic energy-momentum tensor. The term is experimentally detectable under special circumstances at low frequencies [31], but not at optical frequencies, at least not under usual stationary conditions. The Abraham term simply fluctuates out.

The middle term in Eq.(III.1) is the electrostriction term. When seen from within the optically denser medium (the medium with the highest $\varsigma$), the electrostriction force is always compressive. Whether this kind of force is detectable in a static or a stationary case, depends on whether the experiment is able to measure local pressure distributions within the compressed region or not. Moreover, in a dynamic case the velocity of sound is an important factor. If the elastic pressure in the fluid has sufficient time to build up, then the electrostriction force will not be detectable when measuring the gross behavior of a fluid such as the elevation of its surface. Such is usually the case in optics. The time required for the counterbalance to take place, is of the same order of magnitude as the time needed for sound waves to traverse the cross section of the laser beam. For a beam width around *10 μm*, this yields a time scale for counterbalance of the order of *10 ns*. For instance, in the Ashkin-Dziedzic experiment [28] a detailed calculation shows that this time scale was verified; cf. Figure 9 in Ref. [31].

Another point worth mentioning in connection with the electrostriction term is that we have written $(\partial \varepsilon / \partial \rho)_S$ as an adiabatic partial derivative. This seems most natural in optical problems in view of the rapid variations of the field, at least in connection with laser pulses. In many cases it is however legitimate to assume that the medium is nonpolar, so that we need not distinguish between adiabatic and isothermal derivatives. The permittivity depends on the mass density only. Then derivative can be written simply as $d\varepsilon/d\rho$, and is calculable from the Clausius-Mossotti relation. In this way we can write Eq.(III.1) in the following form, when omitting the last term,

$$\vec{f} = -\frac{1}{2}\varepsilon_0 |E|^2 \vec{\nabla}\varsigma + \frac{1}{6}\varepsilon_0 \vec{\nabla}\left[|E|^2 (\varsigma - 1)(\varsigma + 2)\right] \tag{III.2}$$

Finally, we have the first term in Eq.(III.1), which may be called the Abraham-Minkowski force, since it follows equally well from the Abraham and the Minkowski energy-momentum tensors:

$$\vec{f}^{AM} = -\frac{1}{2}\varepsilon_0 |E|^2 \vec{\nabla}\varsigma \tag{III.3}$$

This is the only term that we have to take into account in practice in optics, under usual circumstances. We see that this force is equal to zero in the homogeneous interior of the medium, and acts in the inhomogeneous boundary region only. By integrating the normal component of the Abraham-Minkowski force density across the boundary, we obtain the surface force density which is often evaluated as the jump of the normal component of the electromagnetic Maxwell stress tensor.

In the following we thus focus attention the force term in Eq.(III.3) only. As a parenthetical remark, we note that this expression has shown its applicability in the quantum world also, in connection with the Casimir effect [52]; for recent reviews see [53, 54, 55]. The main expression giving the force between two parallel dielectric slabs because of zero-point oscillations of the electromagnetic field (this is the so-called Lifshitz formula), is actually based upon use of the expression (III.3), combined with the quantum mechanical Green function expression. Modern experiments have actually verified the accuracy of the Lifshitz formula up to an accuracy of a few per cent. Even corrections to the Lifshitz formula, like those arising from finite temperature effects, are at present under active theoretical and experimental study [56].

### *III.2 –Surface Tension and Radiation Forces on a Curved Surface*

To begin with, we assume that there is a curved surface $z = h(x,y,t)$ distinguishing two fluids, a lower fluid (1) and an upper fluid (2). The surface is assumed to have been displaced by the combined effect of gravity, surface tension, and radiation pressure. The equilibrium position of the surface is the $x-y$ plane. Usually, stationary conditions are considered, so that $z = h(x,y)$. However, when at first dealing with the general theory, we shall not limit ourselves to necessarily stationary conditions.

Because of the surface tension coefficient $\sigma$, there will be a normal stress proportional to the mean curvature of the surface:

$$p_1 - p_2 = \sigma \left( \frac{1}{R_1} + \frac{1}{R_2} \right), \qquad (III.4)$$

$R_1$ and $R_2$ being the principal radii of curvature at the surface point considered. If $R_1$ and $R_2$ are positive, $p_1 - p_2 > 0$. This means that the pressure is greater in the medium whose surface is convex. It is useful to have in mind the following general formula for the mean curvature $(1/R_1 + 1/R_2)$:

$$\frac{1}{R_1}+\frac{1}{R_2}=-\frac{-h_{xx}(1+h_y^2)-h_{yy}(1+h_x^2)+2h_{xy}h_xh_y}{(1+h_x^2+h_y^2)^{3/2}},\tag{III.5}$$

where $h_x = \partial h/\partial x$, etc. Also, we note that the unit normal vector $\vec{n}$ to the surface is

$$\vec{n}=(1+h_x^2+h_y^2)^{-1/2}(-h_x,-h_y,1).\tag{III.6}$$

The normal points upwards, from medium 1 to medium 2.

Assume now that there is a monochromatic electromagnetic wave with electric field vector $\overrightarrow{E^{(i)}}(r)e^{-i\omega t}$ incident from below, in the positive z direction. The direction of the incident wave vector $\vec{k}_i$ is thus given by the unit vector:

$$\vec{k}_i=(0,0,1)\tag{III.7}$$

in medium 1. When this wave impinges upon the surface, it becomes separated into a transmitted wave $\overrightarrow{E^{(t)}}(r)$ and a reflected wave $\overrightarrow{E^{(r)}}(r)$, propagating in the directions of $\vec{k}_t$ and $\vec{k}_r$, respectively. We assume, in conformity with usual practice, that the waves can locally be regarded as plane waves and that the surface can locally be regarded as plane. The plane of incidence is formed by the vectors $\vec{k}_i$ and $\vec{n}$; we call the angle of incidence $\theta_i$ and the angle of transmission $\theta_t$. Moreover, we let $\vec{E}_\parallel$ and $\vec{E}_\perp$ be the components of $\vec{E}$ parallel and perpendicular to the plane of incidence, respectively. The expressions for the energy flux transmission coefficients $T_\parallel$ and $T_\perp$ for a plane wave incident upon a boundary surface are (cf. [50],p. 496, or [57]):

$$T_\parallel=\frac{n_2}{n_1}\frac{\cos\theta_t}{\cos\theta_i}\left(\frac{E_\parallel^{(t)}}{E_\parallel^{(i)}}\right)^2=\frac{\sin 2\theta_i \sin 2\theta_t}{\sin^2(\theta_i+\theta_t)\cos^2(\theta_i-\theta_t)},\tag{III.8}$$

$$T_\perp=\frac{n_2}{n_1}\frac{\cos\theta_t}{\cos\theta_i}\left(\frac{E_\perp^{(t)}}{E_\perp^{(i)}}\right)^2=\frac{\sin 2\theta_i \sin 2\theta_t}{\sin^2(\theta_i+\theta_t)}.\tag{III.9}$$

When dealing with an unpolarized radiation field, one usually averages over the two polarizations and represents the transmission coefficient by the single entity

$$\langle T \rangle = \frac{1}{2}(T_\parallel + T_\perp). \tag{III.10}$$

Consider now the electromagnetic surface force density, which we will call $\Pi$. As mentioned above, $\Pi$ can be found by integrating the normal component of the volume force density across the surface boundary layer. From Eq.(III.3) it follows that the surface force acts normal to the surface, and that it is directed towards the optically thinner medium.

We introduce the intensity $I$ of the incident beam,

$$I = \varepsilon n_1 c \langle |E^{(i)}|^2 \rangle, \tag{III.11}$$

and let in the most general case $\alpha$ denotes the angle between $\vec{E^{(i)}}$ and the plane of incidence,

$$E_\parallel^{(i)} = E^{(i)} \cos\alpha, \quad E_\perp^{(i)} = E^{(i)} \sin\alpha. \tag{III.12}$$

Then, we can write the surface force density as

$$\Pi = -\frac{I}{2c} \frac{n_2^2 - n_1^2}{n_2} \frac{\cos\theta_i}{\cos\theta_t} \left[ \left( \sin^2\theta_i + \cos^2\theta_t \right) T_\parallel \cos^2\alpha + T_\perp \sin^2\alpha \right]. \tag{III.13}$$

When $\vec{E^{(i)}} = \vec{E_\parallel^{(i)}}$ or $\vec{E^{(i)}} = \vec{E_\perp^{(i)}}$ (i.e., $\alpha = 0$ or $\alpha = \pi/2$) it is often convenient to express $\Pi$ as

$$\Pi = \frac{n_1 I}{c} \cos^2\theta_i \left( 1 + R - \frac{\tan\theta_i}{\tan\theta_t} T \right), \tag{III.14}$$

where $R = 1 - T$ is the reflection coefficient. This expression also holds in the hydrodynamic nonlinear case. The direction of the force is always evident from the physical situation under study. In connection with the experiments we report on, the upper liquid was always the optically denser one. Thus $n_2 > n_1$, the expression for $\Pi$ is negative, and the vector $\vec{\Pi}$ acts downward, normal to the surface as mentioned above.

The case of normal incidence yields

$$T_\parallel = T_\perp = \frac{4n_1 n_2}{(n_1 + n_2)^2}, \qquad \text{(III.15)}$$

$$\Pi = \frac{n_1 I}{c} \frac{n_1 - n_2}{n_1 + n_2}. \qquad \text{(III.16)}$$

### *III.3 – Cylindrical Symmetry*

Because of calculational convenience as well as relevance for real experiments, we shall henceforth assume cylindrical symmetry, using standard cylinder coordinates $(r, \theta, z)$. There is no variation in the azimuthal direction, $\partial h / \partial \theta = 0$. With the notation $h'(r) = \partial h / \partial r$ we have

$$\cos \theta_i = \frac{1}{\sqrt{1 + h'(r)^2}}, \quad \sin \theta_i = \frac{h'(r)}{\sqrt{1 + h'(r)^2}}. \qquad \text{(III.17)}$$

Together with analogous expressions for $\theta_t$ this can be inserted into Eq.(III.13) to yield

$$\Pi = -\frac{2 n_1 I}{c} \frac{1 - n_{12}}{1 + n_{12}} \psi(h'(r), \alpha), \qquad \text{(III.18)}$$

where $n_{12}$ is the relative refractive index,

$$n_{12} = n_1 / n_2, \qquad \text{(III.19)}$$

and $\psi(h'(r), \alpha)$ is the function

$$\psi(h'(r),\alpha) = \frac{(1+n_{12})^2}{\left[n_{12}+\sqrt{1+(1-n_{12}^2)h'(r)^2}\right]}$$

$$\times \left\{\sin^2\alpha + \frac{1+(3-n_{12}^2)h'(r)^2+(2-n_{12}^2)h'(r)^4}{\left[n_{12}h'(r)^2+\sqrt{1+(1-n_{12}^2)h'(r)^2}\right]}\cos^2\alpha\right\}. \qquad (III.20)$$

When the surface is horizontal, $h'(r)=0$, we have $\psi=1$, and $\Pi$ reduces to the expression (III.16).

A property of the expression (III.20) that facilitates practical calculations is that it is quite insensitive with respect to variations in the polarization angle $\alpha$, especially in the case when $n_{12}$ is close to unity, which is in practice most important as we have noted. Thus if we draw curves for $\Pi(\theta_i)$ versus $\theta_i$ for various input values of $\alpha$ in the whole region $0<\alpha<90°$, we will find that the curves lie close to each other [58]. This is illustrated in Figure III.1 and Figure III.2 for typical experimental conditions used in the following. Represented is the normalized radiation pressure $\Pi(\theta_i)/\Pi(\theta_i=0)$ on a flat interface versus angle of incidence $\theta_i$ when incidence occurs from the optically thin (Figure III.1) and the optically thick (Figure III.2) medium. For comparison, we show in Insets the same curves obtained for the water free surface, where the large index contrast enhances the dependence in polarization For practical calculations involving unpolarized light it is thus legitimate to replace $\psi(h'(r),\alpha)$ by its average with respect to $\alpha$. As $\langle \sin^2\alpha\rangle=\langle \cos^2\alpha\rangle=1/2$, we can then write the surface force density as

$$\Pi = -\frac{2n_1 I}{c}(1-n_{12}^2)\frac{1+(2-n_{12}^2)h'(r)^2+n_{12}h'(r)^2\sqrt{1+(1-n_{12}^2)h'(r)^2}+h'(r)^4}{\left[n_{12}+\sqrt{1+(1-n_{12}^2)h'(r)^2}\right]^2\left[n_{12}h'(r)^2+\sqrt{1+(1-n_{12}^2)h'(r)^2}\right]^2}.$$
(III.21)

This expression is valid also in the case of hydrodynamic nonlinearity. Note again that $\Pi$ is the normally directed force per unit area of the oblique liquid surface.

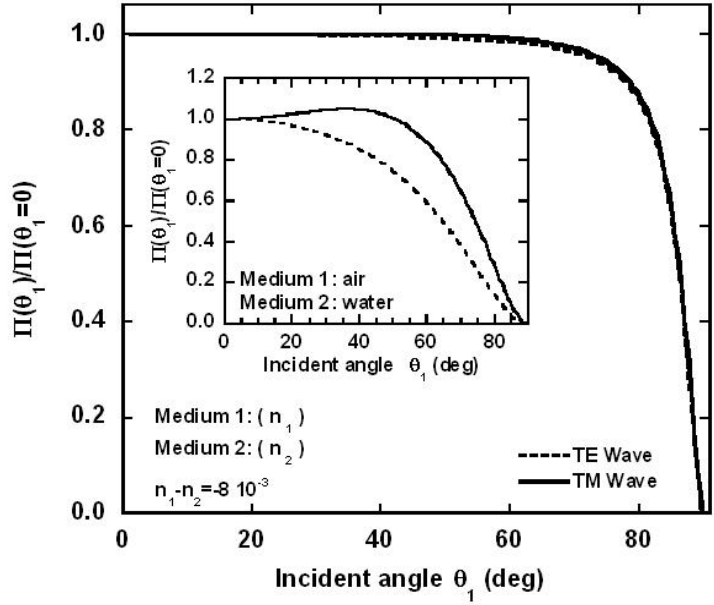

**Figure III.1:** Normalized surface force density on a flat interface versus angle of incidence for both TE and TM polarization when $|n_{12}| \ll 1$, as in the experiments presented below. The light ray is incident from medium 1 (of smallest refractive index) to medium 2. Inset, the same curves are presented for the water/air interface. The ray is incident from air ($n_1 = 1$) to water ($n_2 = 1.33$).

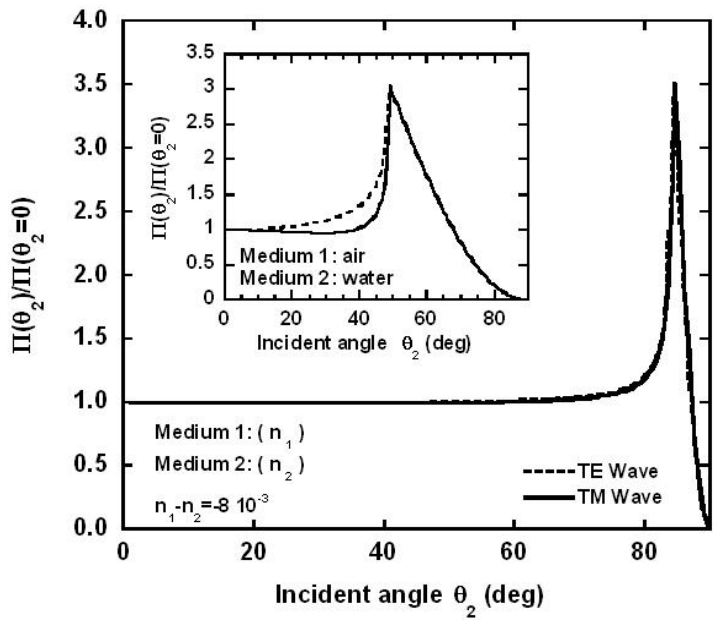

**Figure III.2:** Same as in Figure III.2 but with a reverse direction of the incident ray, i.e. from medium 2 (of largest refractive index) to medium 1 or, from water to air in Inset.

Finally, let us consider the force balance on a curved interface, assuming stationary conditions. When $n_2 > n_1$ the surface tension force which acts upward, has to balance the combined effect of gravity and electromagnetic surface force, which both act downward. When the surface is given as $h = h(r,\theta)$, the mean curvature can be written as

$$\frac{1}{R_1}+\frac{1}{R_2} = -\frac{1}{r}\frac{rh'(r)}{\sqrt{1+h'(r)^2+\frac{1}{r^2}\left(\frac{\partial h}{\partial \theta}\right)^2}} - \frac{1}{r^2}\frac{\partial}{\partial \theta}\frac{(\partial h/\partial \theta)}{\sqrt{1+h'(r)^2+\frac{1}{r^2}\left(\frac{\partial h}{\partial \theta}\right)^2}}, \quad \text{(II.22)}$$

with sign conventions the same as in Eq.(III.5). Thus for azimuthal symmetry,

$$\frac{1}{R_1}+\frac{1}{R_2} = -\frac{1}{r}\frac{d}{dr}\frac{rh'(r)}{\sqrt{1+h'(r)^2}}, \quad \text{(III.23)}$$

and the force balance becomes

$$(\rho_1-\rho_2)gh(r) - \frac{\sigma}{r}\frac{d}{dr}\left(\frac{rh'(r)}{\sqrt{1+h'(r)^2}}\right) = \Pi(r). \quad \text{(III.24)}$$

This equation follows from considering the equilibrium of a curved surface having unit base area.

Which expression to insert for $\Pi(r)$ in Eq. (III.24), depends on the physical circumstances. Thus in the case of an unpolarized laser beam, we may use either the expression (III.14) with $R = \langle R \rangle$; $T = \langle T \rangle$, or alternatively use the expression (III.21). As already noted, there is no restriction imposed here on the magnitude of the slope of the surface.

# IV – Experimental Section

We have seen in Sec. II that deformations induced by a continuous laser beam on classical interfaces were nanometric and required high intensity laser pulses to observe a microscopic height. Thus, direct visualization of micrometric deformations induced by a c.w. Argon ion laser obviously necessitates very soft interfaces. To do so, we choose to work with near-critical interfaces because surface tension vanishes close to a critical point as $\sigma = \sigma_0 |T/T_C - 1|^{1.26}$, where $T$ and $T_C$ are respectively the actual and the critical temperature. Moreover, at a given distance to the critical temperature, supramolecular liquids lead to even smaller surface tension because the amplitude $\sigma_0$ is inversely proportional to the square of the characteristic molecular length scale $\xi_0$, according to renormalization group theory. Consequently, in order to enhance radiation pressure effects on fluid interfaces, we performed experiments in near critical phase separated micellar phase of microemulsions. As they belong to the same universality class ($d = 3$, $n = 1$) of the Ising model as most of isotropic liquids, our particular choice will be of general meaning and can be transposed to any interface separating liquids belonging to the same universality class.

## IV.1 – Properties of the Micellar Phase of Microemulsion used

A polar hydrophilic head and a hydrophobic tail generally compose amphiphile molecules. This peculiarity allows them to self-assemble in solution in various forms and architectures. Micellar phases of microemulsion correspond to one of these thermodynamically stable structures. They are obtained by mixing at least three constituents: water, oil, and a surfactant. A fourth constituent (generally an alcohol called co-surfactant) is nevertheless added to increase the mixture stability. When the volume fraction of added water is of the order of a few percents, surfactant molecules adsorb at the water/oil interface to form a spherical shell around water nano-droplets. This assembling, constituted by surfactant-coated water nano-droplets suspended in oil, is called micellar phase of microemulsion. Consequently, this mixture can be assimilated to a binary mixture composed of a solute (the micelles) and a solvent (the oil). In our experiments, we used a mixture of sodium dodecyl sulfate (surfactant), water, toluene (oil) and n-buthanol-1 (alcohol). Its mass composition in weight fraction is *4 %* SDS, *9 %* water, *70 %* toluene, and *17 %* n-butanol-1. For this composition, the micelle radius is $\xi_0 = 40 \pm 2\ Å$, i.e. one order of magnitude larger than classical molecular length scale, and small enough to still be transparent in the visible optical wavelength window. This composition was also chosen so as to be critical at a temperature $T_C = 35\ °C$ [59]. As illustrated in Figure IV.1, for a temperature $T > T_C$, the mixture phase-separates in two micellar phases of different concentration $\Phi$. According to renormalization group theory, the amplitudes $\sigma_0$ and $\xi_0$ are related by a universal ratio $R^+$ defined as $R^+ = \left(\sigma_0 \xi_0^2 / k_B T_C\right) = 0.39$ [60]. This means that surface tension in micellar systems is weak compared to liquid-vapor interfaces. For example, for our particular mixture at *1.5 K* from the critical point, $\sigma$ is one million times smaller than that of the water free surface ($\sigma = 72\ mJ.m^{-2}$). This weakness clearly illustrates the reason why we can expect large interface deformations using c.w. lasers.

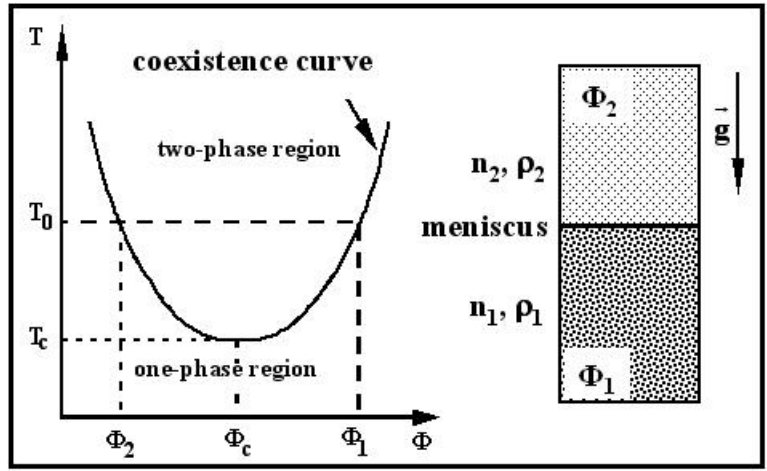

**Figure IV.1:** Left: schematic phase diagram of the micellar phase of microemulsion used. *T* is temperature and $\Phi$ is the volume fraction of micelles. Right: experimental configuration for a temperature $T_0 > T_C$.

Another advantage of near-critical systems is the possibility to continuously tune fluid properties by simply varying temperature. In summary, one has:

Coexistence curve $\Delta\Phi = \Delta\Phi_0 \left(\dfrac{T - T_C}{T_C}\right)^\beta$ with $\beta = 0.325$ and $\Delta\Phi_0 = 1.458$

Density contrast: $\Delta\rho = \Delta\rho_0 \left(\dfrac{T-T_C}{T_C}\right)^\beta$ with $\Delta\rho_0 = \left(\dfrac{\partial\rho}{\partial\Phi}\right)\Delta\Phi_0 = 284\ kg.m^{-3}$

Refractive index contrast: $\Delta n = \Delta n_0 \left(\dfrac{T-T_C}{T_C}\right)^\beta$ with $\Delta n_0 = \left(\dfrac{\partial n}{\partial\rho}\right)\Delta\rho_0$ and $\left(\dfrac{\partial n}{\partial\rho}\right) = -1.22\times 10^{-4}\ m^3.kg^{-1}$

Surface tension: $\sigma = \sigma_0 \left(\dfrac{T-T_C}{T_C}\right)^{2\nu}$ with $\nu = 0.63$ and $\sigma_0 = 10^{-4}\ J.m^{-2}$

As these quantities can be varied over significant range, they illustrate the usefulness of near critical fluid interfaces for investigating the influence of fluid properties on interface deformation using a single sample and low power lasers. By combining these critical amplitudes, we can also predict the critical behavior of more complex quantities that appear when dealing with interface deformations, such as the capillary length $l_C$, defined as:

$$l_C = \sqrt{\dfrac{\sigma}{\Delta\rho g}} = l_{C_0}\left(\dfrac{T-T_C}{T_C}\right)^{\nu-\beta/2} \text{ with } l_{C_0} = \sqrt{\dfrac{\sigma_0}{\Delta\rho_0 g}}$$

Finally, in order to analyze radiation pressure effect, thermal heating should not disturb interface deformation. This condition is fulfilled for our micellar phases since the optical absorption is $\alpha_a = 3.10^{-4}\ cm^{-1}$.

*IV.2 – Experimental setup*

The experimental setup is presented in Figure IV.2. As beam intensity depends on both power and beam radius, the main point is to form a beam waist $\omega_0$ in the cell C on the meniscus of the phase-separated mixture and to vary its size. To do so, we use the lens $L_1$ to form a first intermediate waist. The variation of the beam waist $\omega_0$ is performed by moving the prism Pr to increase the optical path between $L_1$ and the focusing long working distance microscope objectives, either $O_1$ or $O_2$ ($10\times$). As the beam waist position varies within the cell, C is mounted on translation stages to locate $\omega_0$ on the interface. Moreover, since interface deformation should not depend on the beam propagation direction, our setup allows for ascending and descending beams by using the half-wave plate $\lambda/2$ and the beam splitter BS. Interface deformations are induced by the radiation pressure of a c.w. Ar+ laser (wavelength in vacuum $\lambda_0 = 0.514\ \mu m$) in the TEM$_{00}$ mode. Using this procedure accessible values are $\omega_0 = 4.8 - 32.1\ \mu m$ (resp. $\omega_0 = 3.5 - 11.3\ \mu m$) for ascending (resp. descending) beams. Considering weakly focalized beams, we assume negligible beam divergence around the focus and write the intensity profile as:

$$I(r,z) \approx I(r) = \frac{2P}{\pi \omega_0^2} exp\left(-\frac{2r^2}{\omega_0^2}\right)$$

where $P$ is the incident beam power. In fact, this approximation is correct as far as $\lambda_0 h / (n\pi\omega_0^2) < 1$, where $h$ and $n$ are respectively the height of the deformation and the index of refraction. On the other hand, observations are performed transversally using a white light source for illumination and a microscope for imaging on a C.C.D. video camera coupled to a frame grabber. A spectral filter is also placed between the microscope and the C.C.D. to eliminate residual laser light scattered by the micellar phases.

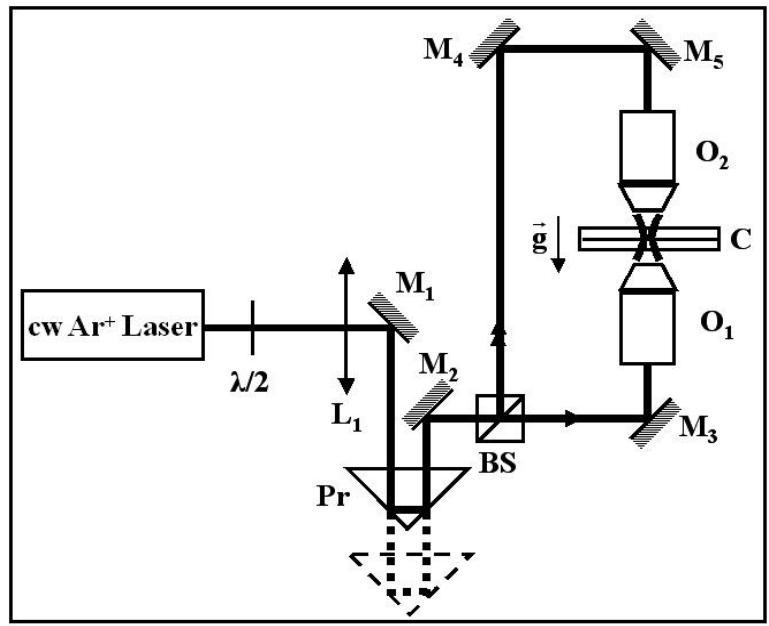

**Figure IV.2:** Experimental setup. $\lambda/2$ : $\lambda/2$ plate, $L_1$: lens, $M_{i=1,5}$: mirrors, $Pr$: prism; $BS$: beam splitter; $O_{i=1,2}$ microscope lens 10×; $C$: experimental cell.

## V – Interface Bending in the Linear Regime in Deformation [61]

While radiation pressure effects on interfaces are usually weak, in the preceding section we have seen how to significantly enhance laser-induced interface deformations. Consequently it becomes possible to directly visualize interface bending. This section is devoted to what we call the linear regime, i.e. the regime for which the bending slope is weak. While it corresponds to the most classical situation, the following section will present results when deformations become highly nonlinear.

### V.1 – Stationary interface deformations

Figure V.1 gives two typical examples of interface bending observed in different conditions for an ascending beam (Figure V.1a: $T - T_C = 8\ K$, $\omega_0 = 7.5\ \mu m$, and Figure V.1b: $T - T_C = 2\ K$, $\omega_0 = 14.6\ \mu m$). As the refractive index $n_1$ of the incident fluid phase is lower than that of the upper phase, interface deformation occurs downward as expected. Its height also increases versus beam power. Moreover, by comparing both series, we see that its shape depends significantly on the beam waist. Temperature dependent effects are less obvious, but they will be evidenced in the following. To explain the shape of these deformations, let us consider a flat fluid interface at rest and of infinite extension. The general expression governing the steady state interface shape $h(r)$ under radiation pressure effects is given in cylindrical coordinates by:

$$(\rho_1 - \rho_2)gh(r) - \sigma\kappa(r) = \Pi(r), \tag{V.1}$$

where $\kappa(r)$ is the interface curvature. Eq. (V.1) shows that the height of the deformation results from the compensation of radiation pressure $\Pi(r)$ by both buoyancy $(\rho_1 - \rho_2)gh(r)$ and Laplace force $-\sigma\kappa(r)$, where $\kappa(r)$ is expressed as

$$\kappa(r) = \frac{1}{r}\frac{d}{dr}\left(\frac{rh'(r)}{\sqrt{1+h'(r)^2}}\right). \tag{V.2}$$

Considering one single reflection/refraction, the surface pressure $\Pi(r)$ caused by the incident beam is given by the formula [62, or Eq. (III.14)]:

$$\Pi(r) = \frac{n_i I}{c}\cos^2\theta_i\left(1 + R - \frac{\tan\theta_i}{\tan\theta_t}T\right). \tag{V.3}$$

The index $i$ and $t$ respectively refer to incident and transmitted ray, and $R$ and $T$ are the Fresnel coefficients of reflection and transmission that satisfy $R + T = 1$ [63]. However, for weak deformations, one can use the expression of $R$ and $T$ at normal incidence, and linearize curvature. In these conditions Eq. (V.1) reduces to

$$(\rho_1 - \rho_2)gh(r) - \sigma\Delta_r h(r) = \frac{2n_1}{c}\left(\frac{n_1 - n_2}{n_1 + n_2}\right)I(r). \tag{V.4}$$

Eq. (V.4) describes the shape of the induced deformation in the linear regime in deformation. Before solving this equation, we perform a dimensional analysis of the different terms involved in order to point out the relevant physical parameters.

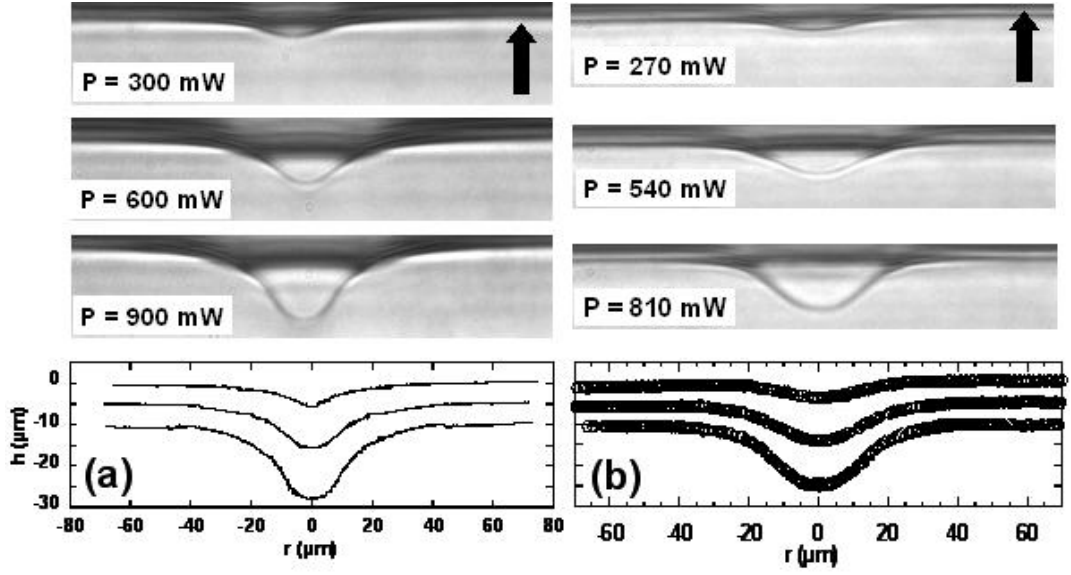

**Figure V.1:** Variation of the interface deformation by optical radiation pressure for increasing beam powers. Black arrows indicate the direction of beam propagation. The fits (empty circles) are performed according to Eq. (V.14). Control parameters are (a) $T - T_C = 8\ K$, $\omega_0 = 7.5\ \mu m$, and (b) $T - T_C = 2\ K$, $\omega_0 = 14.6\ \mu m$ ).

### Capillary length $l_C$ and Optical Bond Number

As radiation pressure is balanced by $(\rho_1 - \rho_2)gh(r)$ and $-\sigma\Delta_r h(r)$, buoyancy and curvature can as well compete together and their comparison leads to introduce a new length scale, the capillary length $l_C$. If $R$ denotes the typical width associated to the radial extension of the induced deformation, $\Delta_r h(r)$ scales as $h/R^2$. Consequently, the ratio between buoyancy and surface effects leads to define a Bond Number $B$ as:

$$B = \frac{\Delta\rho g h}{\sigma h/R^2} = \left(\frac{R}{l_C}\right)^2, \tag{V.5}$$

where $l_C = \sqrt{\sigma/\Delta\rho g}$ is the capillary length. This is an intrinsic medium-dependent length scale. To compare the relative importance of buoyancy versus surface tension effects on interface deformation, one has to compare the characteristic extension $R$ to $l_C$. For $R \ll l_C$, i.e. at small length scales, deformation is dominated by surface tension while in the opposite case $R \gg l_C$ radiation pressure will mainly compete with buoyancy. As deformations are due to the radiation pressure of a laser beam of beam waist $\omega_0$, we can assume $R \approx \omega_0$ and define an "Optical" Bond Number $Bo$ as:

$$Bo = \left(\frac{\omega_0}{l_C}\right)^2. \tag{V.6}$$

This means that the induced deformation can be viewed as a sort of virtual particle of radius $\omega_0$. Moreover, as $l_C$ is a characteristic medium-dependent length scale that only depends on the vicinity to the critical point, and $\omega_0$ is a second length scale of optical nature that can be tune externally and independently to $l_C$, existence of a relevant Bond Number gives us the possibility to scan interface deformation driven by radiation pressure from the low to the large $Bo$ regime.

### V.2 – Asymptotic regimes and universal behavior

#### Height of deformations is the weak optical Bond number regime

The $Bo \ll 1$ regime corresponds to the case of usual interfaces [28, 35, 42], where capillary waves are mainly driven by surface tension. Eq. (V.4) becomes:

$$-\sigma \Delta_r h(r) \approx \frac{2n_1}{c}\left(\frac{n_1 - n_2}{n_1 + n_2}\right) I(r). \tag{V.7}$$

To solve Eq. (V.7), we assume the following boundary condition, $h = 0$ for $r = \omega_{bc}/\sqrt{2}$ and $\omega_{bc} \gg \omega_0$, to remove the classical logarithmic divergence at infinity of Laplacian equation in cylindrical coordinates. We find:

$$h(r) = \frac{2n_1}{c}\left(\frac{n_1-n_2}{n_1+n_2}\right)\frac{P}{4\pi\sigma}\left[E_1\left(\frac{\omega_{bc}^2}{\omega_0^2}\right) - E_1\left(\frac{2r^2}{\omega_0^2}\right) - ln\left(\frac{2r^2}{\omega_{bc}^2}\right)\right], \quad \text{(V.8)}$$

where, $E_1$ is the 1-argument exponential function. Using, $E_1(x) \approx -ln(\gamma x) + x$ close to beam axis, where $\gamma = 1.781$ is the Euler constant, the height on the induced deformation is:

$$h(r=0) = \frac{2n_1}{c}\left(\frac{n_1-n_2}{n_1+n_2}\right)\frac{P}{4\pi\sigma} ln\left(\gamma\frac{\omega_{bc}^2}{\omega_0^2}\right). \quad \text{(V.9)}$$

Note that the height of the deformation in the $Bo \ll 1$ regime is proportional to the beam power $P$ instead of axial beam intensity $I_0 = 2P/\pi\omega_0^2$. This means that the $Bo \ll 1$ regime corresponds to a nonlocal regime in deformation. This peculiarity is easily understood in Fourier space. If $I(q)$ denotes the Fourier component of the beam intensity associated to the mode $q$, then $h(q) \propto I(q)/q^2$. This nonlocal character is then illustrated by the divergence of the $q=0$ mode. The beam waist does not influence deformation anymore. A larger length scale should control it instead. As we investigate the $Bo \ll 1$ regime, the other length scale is necessarily the capillary length. Considering that $\Phi_1$ and $\Phi_2$ are coexisting phases of close composition due to the vicinity of the critical point, we assume $n_1 \approx n_2$ and $\Delta n \approx (\partial n/\partial \rho)\Delta\rho$ and, we rewrite the expression of $h(r=0)$ as:

$$h(r=0) = \left(\frac{\partial n}{\partial \rho}\right)\frac{1}{4\pi cg} ln\left(\frac{8l_C^2}{\gamma\omega_0^2}\right)\frac{P}{l_C^2}, \quad \text{(V.10)}$$

where $\omega_{bc} = 2\sqrt{2}l_C/\gamma$ (see below). Eq. (V.10) shows that the capillary length is, as expected the characteristic length that drive deformation at low optical Bond number. Consequently, data obtained in different conditions should be plotted versus $P/l_C^2$ to point out a single-scaled behavior, rather than $P/\omega_0^2$. The prefactor $(\partial n/\partial \rho)$ is a constant, which can be obtained by the Claussius-Mossotti relation because the index and the density contrast have the same critical behavior.

To investigate the $Bo \ll 1$ regime, experiment should be performed "far" from the critical point with focused beams. Figure V.2 confirms the expected scaling for $\omega_0 = 7.5\ \mu m$ and five large values of $(T - T_C)$. The dispersion in slope observed when the height of the deformation is represented versus beam intensity $P/\omega_0^2$ totally disappears for all $(T - T_C)$ and $\omega_0$ when $P/l_C^2$ is used.

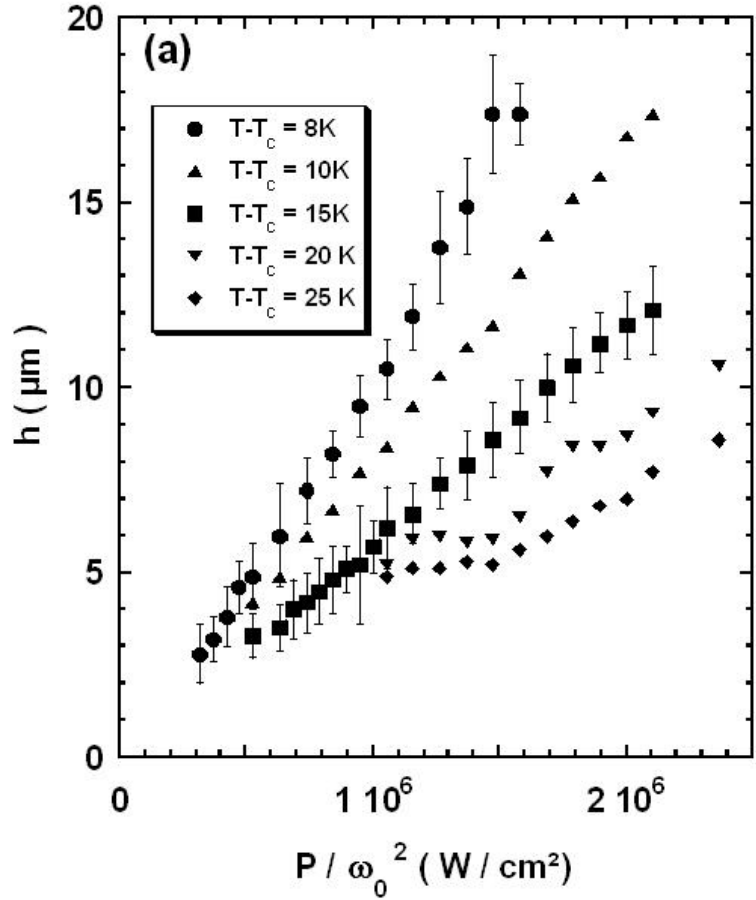

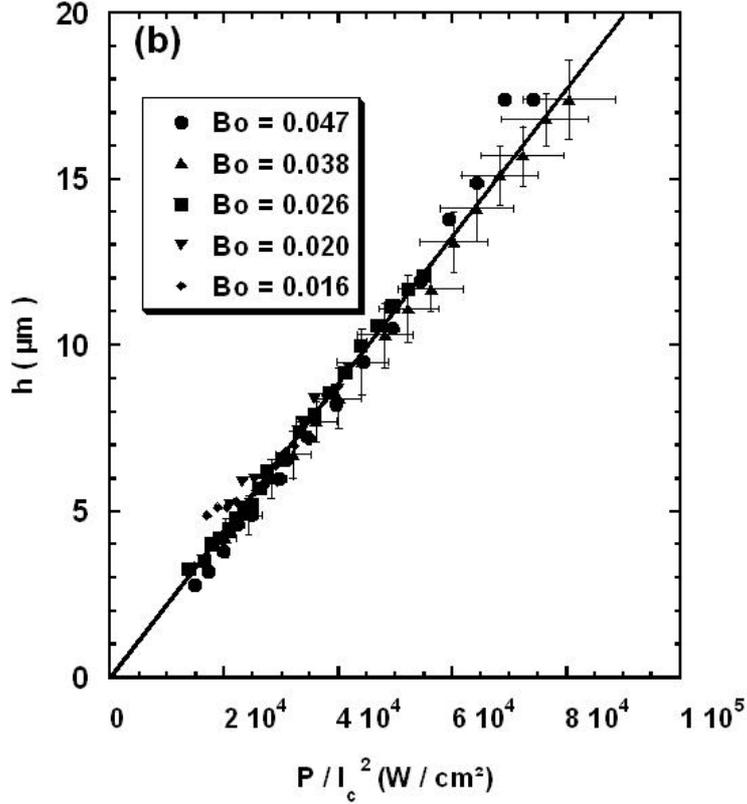

**Figure V.2:** Experimental variation of the height of the deformation at low optical Bond numbers. (a) Variations when plotted versus $I_0 \approx P/\omega_0^2$. (b) Same data represented versus the scaling predicted by Eq. (V.10). The beam waist value is $\omega_0 = 6.3~\mu m$.

Moreover, Figure V.3 generalizes this result to three sets of experiments performed for three different beam waists. As expected from Eq. (V.10), very close slopes, due to the logarithmic correction, are retrieved for the three experiments Consequently, these experiments firmly confirm already existing disparate results by describing interface deformation over large variation in optical and thermodynamic length scales. It also illustrates the strength of critical fluids as well as the associated thermodynamic universality. Indeed, if we have been able to illustrate scaling for the first time, this is because we used fluid interfaces whose properties can be continuously tuned. However, the $Bo \ll 1$ regime corresponds in fact to the case of usual interfaces for which buoyancy is generally neglected. Another peculiarity of our system is also to allow the investigation of the opposite $Bo \gg 1$ regime, where deformations are stabilzed by gravity.

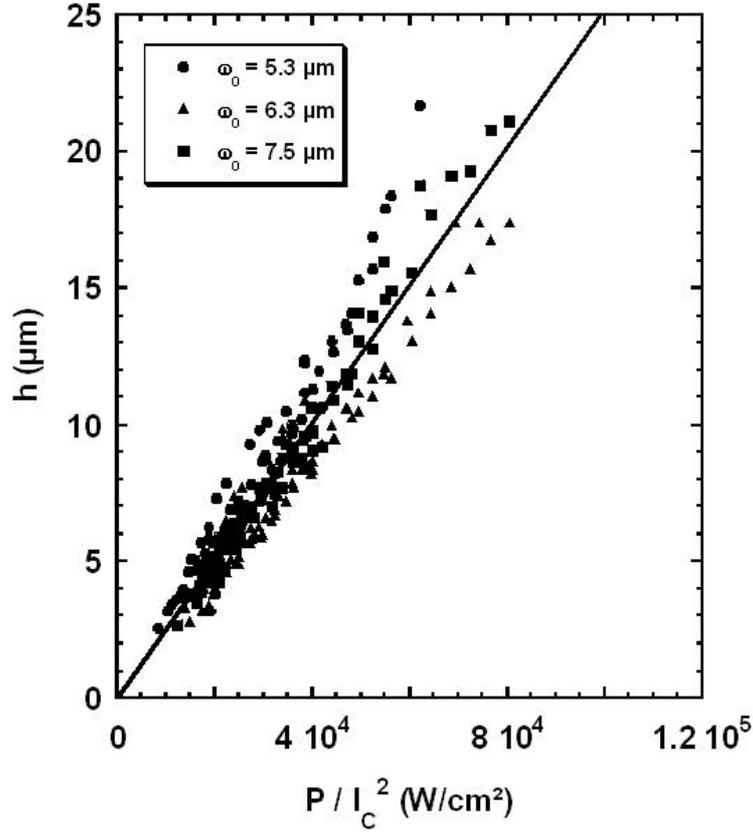

**Figure V.3:** Generalization of the scaling presented in Figure V.2b for $Bo \ll 1$ to data sets performed for three different beam waists.

*Height of deformations is the large optical Bond number regime*

When $Bo \gg 1$, the systems reaches a unusual regime where surface tension effects become negligible compared to buoyancy. In this case, according to Eq. (V.4), the height of the induced deformation is described by:

$$(\rho_1 - \rho_2)gh(r) = \frac{2n_1}{c}\left(\frac{n_1 - n_2}{n_1 + n_2}\right)I(r). \qquad (V.11)$$

Contrary to the $Bo \ll 1$ regime, the height of the deformation becomes proportional to beam intensity. The response of the interface to the optical excitation becomes local; in Fourier space one has $h(q) \propto I(q)$. Using the same approximations as before we find:

$$h(r=0) = \left(\frac{\partial n}{\partial \rho}\right)\frac{1}{cg}\frac{2P}{\pi\omega_0^2} \qquad (V.12)$$

Consequently, data obtained in different temperature conditions should be plotted versus $P/\omega_0^2$ to point out a single-scaled behavior. The prefactor $(\partial n/\partial \rho)$ is still constant, due to the fact that index and density contrast have the same critical behavior.

To investigate the $Bo \gg 1$ regime, experiment should now be performed very close to the critical point with large beams. Results are presented in Figure V.4 for $(T - T_C) = 2\ K$. At first, we observe a linear variation of the height of the deformation versus $P/\omega_0^2$, i.e. beam intensity.

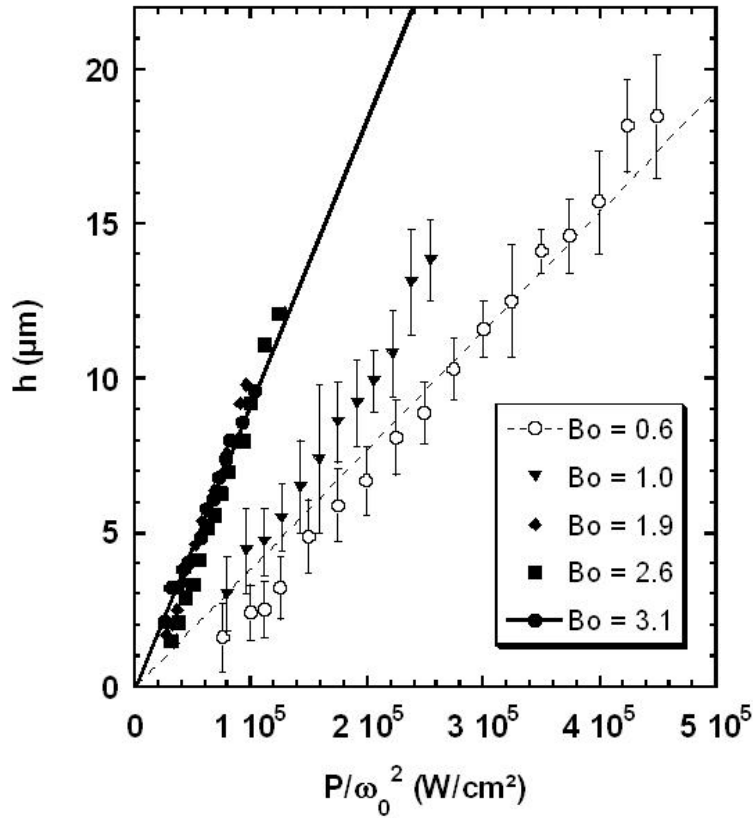

**Figure V.4:** Experimental variation of the height of the deformation for optical Bond numbers increasing from the intermediate to the large Bo regime. The scaling predicted by Eq. (V.12) for $Bo \gg 1$ is illustrated by the limiting slope of $h(r=0)$ versus $P/\omega_0^2$ reached

for the two largest Bo. Control parameters are $(T-T_C) = 2\ K$ and $\omega_0 = 14.6\ \mu m$ (○), $\omega_0 = 18.3\ \mu m$ (▼), $\omega_0 = 25.3\ \mu m$ (◆), $\omega_0 = 29.3\ \mu m$ (■), $\omega_0 = 32.1\ \mu m$ (●).

Moreover, although slopes of $h(r=0)$ versus $P/\omega_0^2$ are well separated at small Bond numbers, they reach progressively a finite common value for $Bo > 1$. Finally, we have represented in Figure V.5 the experimental data obtained at $(T - T_C) = 1.5\ K$ and $(T - T_C) = 2\ K$, and corresponding to the $Bo > 1$ regime. As expected, measurements show a single-scaled behavior, which demonstrates that the slope of $h(r=0)$ versus $P/\omega_0^2$ does not vary with temperature. Consequently, the extremely weak surface tension reached close to the critical point allows, for the first time, investigation of the local regime in deformation were heights become proportional to electromagnetic intensity.

$Bo \ll 1$ and $Bo \gg 1$ case correspond to asymptotic regimes were the involved mechanisms can clearly be evidenced and illustrated. In the following, we will show that the height of the deformations can be described by a universal function of $Bo$, whatever the Bond number is.

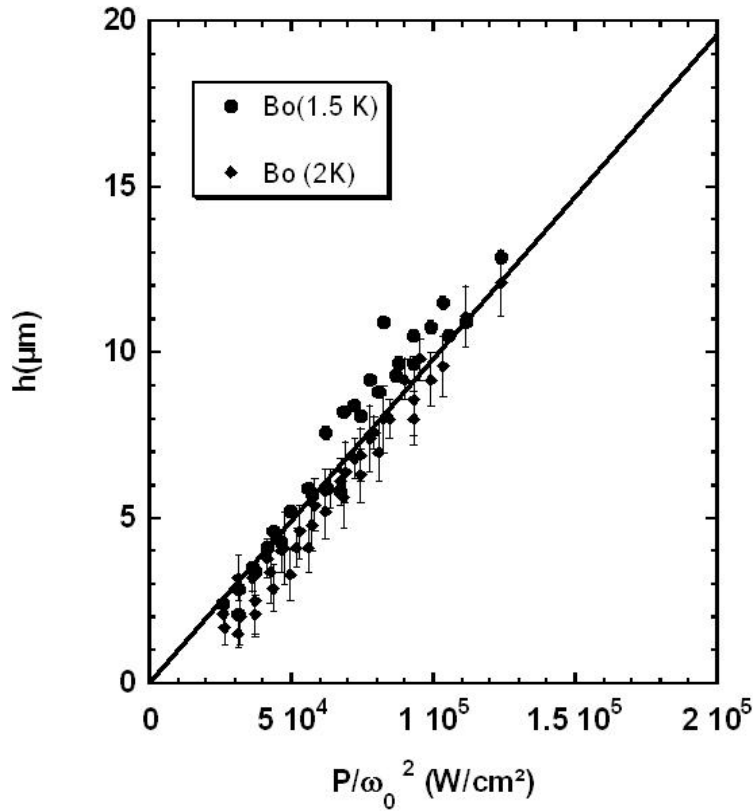

**Figure V.5:** Generalization of the scaling presented in Figure V.4 for $Bo \gg 1$ to data sets performed at two different values of $(T - T_C)$: $(T - T_C) = 1.5\ K$ (●) and $(T - T_C) = 2\ K$ (◆).

*Scaling and universality in interface deformation*

To find a general scaling the deformation height versus optical excitation, we need to integrate Eq. (V.4). As the beam intensity is of cylindrical symmetry, we use the Fourier – Bessel transform defined as:

$$h(r) = \int_0^\infty \tilde{h}(k) J_0(kr) k\, dk, \tag{V.13}$$

where $J_0$ is the is the $0^{th}$ order Bessel $J$ function. This transform is particularly adapted to problems involving Gaussian functions because $exp(-2r^2/\omega_0^2)$ can be decomposed in Fourier-Bessel modes. Using this decomposition, the general solution of Eq. (V.4) is:

$$h(r) = \frac{n_1}{c}\left(\frac{n_1 - n_2}{n_1 + n_2}\right)\frac{P}{\pi \Delta \rho g}\int_0^\infty J_0(kr)\frac{exp(-\omega_0^2 k^2/8)}{1 + k^2 l_C^2} k\, dk. \tag{V.14}$$

Eq. (V.14) was used to compare prediction with the whole profile of stationary deformations measured experimentally (Figure V.1). The nice feature with this integral solution is that we can extract an analytical expression for the height of the deformation $h(r = 0)$. We find:

$$h(r = 0) = h(r = 0)\big|_{Bo \gg 1} \times F(Bo) \tag{V.15}$$

where $h(r = 0)\big|_{Bo \gg 1} = (\partial n/\partial \rho) I(r = 0)/cg$ corresponds to the height predicted for the $Bo \gg 1$ regime and $F(Bo)$ is a universal function of the $Bo$ number given by:

$$F(Bo) = \frac{Bo}{8} exp\left(\frac{Bo}{8}\right) E_I\left(\frac{Bo}{8}\right). \qquad (V.16)$$

For $Bo \gg 1$ its development is straightforward as $F(Bo)|_{Bo \gg 1} \to 1$. On the other hand the development for the $Bo \ll 1$ regime is $F(Bo)|_{Bo \ll 1} = -Bo/8 \ln(\gamma Bo/8) + 0(Bo)$ which leads to the expression of the boundary condition $\omega_{bc} = 2\sqrt{2} l_C / \gamma$ used in Eq. (V.10).

To illustrate universality in interface deformation driven by the radiation pressure of a laser wave, we represent in Figure V.6 the full set of experiments by plotting $H(Bo) = h(r=0)/h(r=0)_{Bo \gg 1}|$ versus $Bo$. The full line represents the universal function $F(Bo)$. Note that there is no adjustable parameter. The Inset shows the same data in log-log plot. Agreement is observed over almost three orders of magnitude in Bond number. Its universal character is also underlined by the representation of a data point measured at the water free surface in Ref. [42].

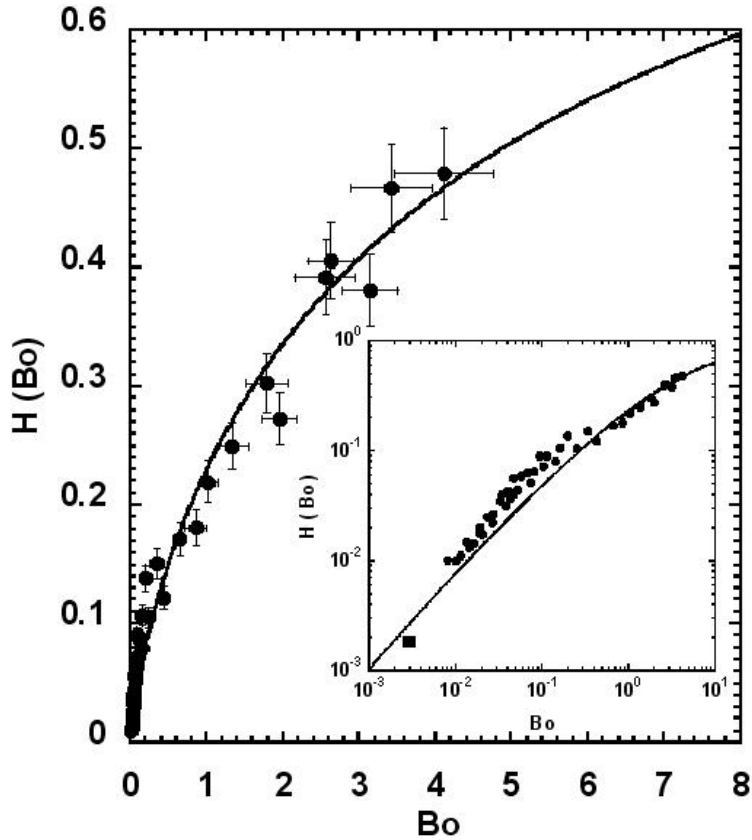

**Figure V.6:** Experimental variation of the dimensionless height of the deformation $H(Bo)$ versus the optical Bond number. The solid line represents the universal scaling function $F(Bo)$ given in Eq. (V.16). A log-log plot overview is presented in the Inset, which also shows the location of the datum (empty circle) given in Ref. [42] for the water/air free surface at room temperature.

Consequently, Figure V.6 can be viewed as a sort of abacus. Knowing the fluid properties, one can predict the deformation versus beam characteristics and conversely, any measurement of the height of the deformation can be used to measure the surface tension in a contact less way. Moreover, as illustrated in Section II, the method can easily be extended to characterization of visco-elastic media, such as biological cells.

*V.3 – Up/down symmetry*

We demonstrate in Section II that interface deformation by the radiation pressure does not depend on direction of beam propagation because photon momentum varies linearly with index of refraction. However, in calculating the Fresnel coefficient of transmission and reflection, we have to take into account the fact that the incident medium changed; using our notations, for a descending beam incidence corresponds to medium 2. The behavior of the interface is then described by:

$$(\rho_1 - \rho_2)gh(r) - \sigma \Delta_r h(r) = \frac{2n_2}{c}\left(\frac{n_1 - n_2}{n_1 + n_2}\right)I(r). \qquad (V.17)$$

This description is totally analogous to that given for an upward beam (Eq. (V.4)), except that a $n_2$ associated to incidence replaces $n_1$ in the amplitude factor of the radiation pressure. As we wanted to verify this prediction, we inverted the direction of propagation by turning the $\lambda/2$ plate (Figure IV.2). We performed a new set of experiments. Two examples are presented in Figure V.7 for the two extreme temperatures of the investigated range. As expected, interface still deforms towards the medium of smaller refractive index. The interface shape is also very similar to those observed with an upward excitation.

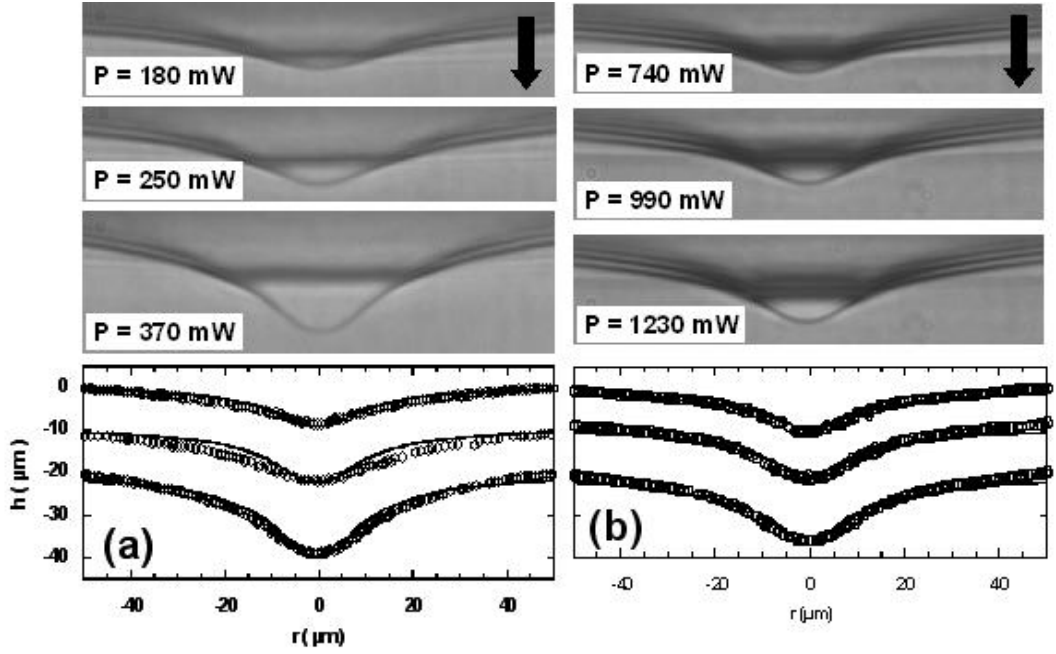

**Figure V.7:** Variation of the interface deformation by optical radiation pressure for increasing beam powers. Black arrows indicate the direction of beam propagation. The empty circles represent the fits. Control parameters are (a) $T - T_C = 3\ K$, $\omega_0 = 7.4\ \mu m$, and (b) $T - T_C = 18\ K$, $\omega_0 = 4.3\ \mu m$).

As the theoretical profiles of deformations induced by upward and downward beams are almost identical due to the fact that $n_1 \approx n_2$ for near-critical interfaces, we should retrieve the same universal behavior in both cases. Results are presented in Figure V.8. We observe a nice superposition of upward and downward data. As far as deformations belong to the linear regime, the predicted scaling law is then valid whatever the beam direction is.

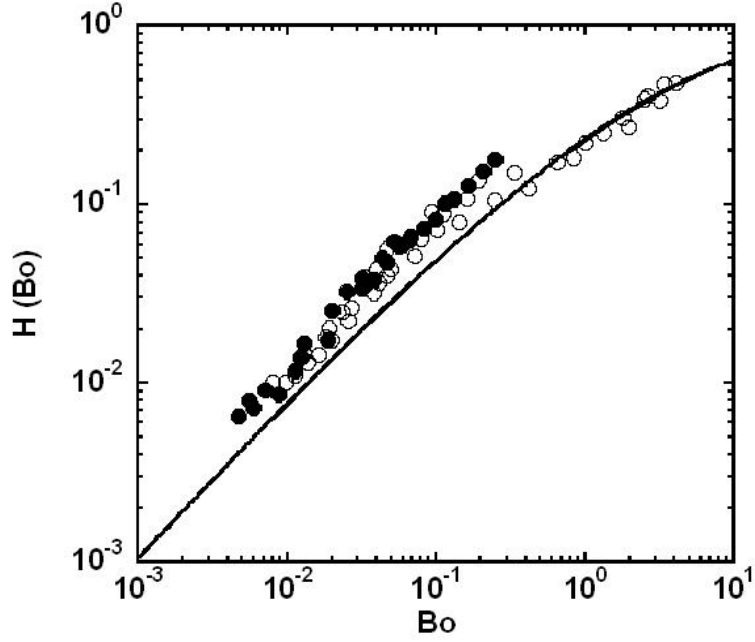

**Figure V.8:** log-log plot of the experimental variation of the dimensionless height of the deformation $H(Bo)$ versus the optical Bond number for both upward (○) and downward (●) beam excitation. The solid line represents the universal scaling function $F(Bo)$ given in Eq. (V.16).

### V.4 – Application to adaptative lensing [64]

The generally weak amplitude (nanometric scale) of the interface deformation induced by radiation pressure of c.w. laser on classical interfaces makes them difficult to characterize, except by using the associate lensing. Indeed, the deformation corresponds to a curved interface between two media of different refractive indices, and can thus be assimilated to a "soft" lens. In the paraxial approximation, the focal distance $f$ is proportional to curvature [65]:

$$\frac{1}{f} = \frac{n_2 - n_1}{2n_2}\kappa \approx \frac{n_2 - n_1}{2n_2}(\Delta h)\Big|_{r=0}. \tag{V.18}$$

The expression of $(\Delta h)\Big|_{r=0}$ can be calculated from Eq. (V.4) and Eq. (V.18) becomes:

$$\frac{1}{f} = \frac{(n_2 - n_1)^2}{2n_2} \frac{I(r=0)}{c\sigma} \left[1 - F(Bo)\right]. \qquad (V.19)$$

For classical fluid interfaces, $Bo \ll 1$ and then $F(Bo) \ll 1$. Then Eq. (V.19) reduces to:

$$\left.\frac{1}{f}\right|_{Bo \ll 1} = \frac{(n_2 - n_1)^2}{2n_2} \frac{I(r=0)}{c\sigma}. \qquad (V.20)$$

Eq. (20) shows that the focal distance associated to a deformation is directly proportional to surface tension. Even for weak deformations, one can measure the associated focal distance by analyzing in far field the intensity profile modification of a second probe beam. This is the method used by Sakai and coworkers [42, 43] to measure surface tensions. It is particularly efficient for very weak surface tension as far as one can calibrate the setup with an interface of known surface tension. As in our experiments deformations are microscopic, we can easily revert the method and measure actual lensing driven by radiation pressure.

Considering Eq. (V.20), we expect the following critical behavior for the focal distance at fixed beam intensity:

$$f \propto (T - T_C)^{2(\nu - \beta)} \qquad (V.21)$$

Figure V.9 illustrates this behavior for $1.5 \leq T - T_C \leq 18\ K$; lines correspond to predictions. Agreement is observed with the expected critical exponent. Moreover, the interesting point here is the tuning of the focal distance by temperature. For example, at a given beam intensity, it is possible to control the focal distance over a large variation ($0.5 \leq f \leq 2.5\ mm$) just by changing temperature.

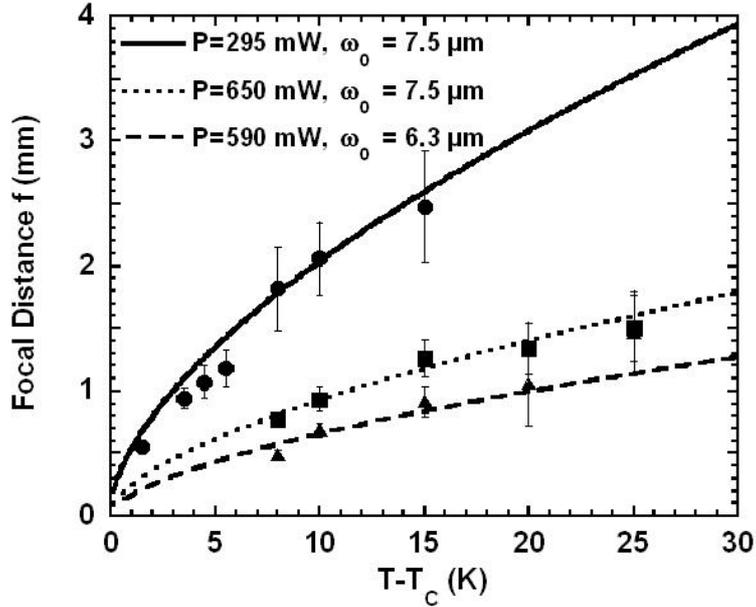

**Figure V.9:** Variation of focal length induced by optical bending versus temperature. The curves represent predictions from Eq. (V.21).

On the other hand, once the range of variation is fixed by temperature, further adjustment can be realized by varying beam intensity. As predicted in Eq. (V.20) for the $Bo \ll 1$ regime, $f$ is inversely proportional to beam intensity. Figure V.10 illustrates this behavior. A power law fit retrieves the expected hyperbolic behavior.

One can thus consider the deformation of a fluid interface as a lens of variable focal distance. By adjusting temperature and beam parameters, we are able to reversibly adapt $f$ over almost two orders of magnitude, typically from *0.5* to *50 mm*. Even if the most reliable adaptative lenses are actually based on electro-wetting (see [66] for a review), soft lenses induced by the optical radiation pressure present many advantages. First, they are completely optically driven. Second, the adaptation range is important, and several setups that use such a liquid lens were devised [67]. One might nevertheless wonder at which level of development our type of sensitive media can be confidently used.

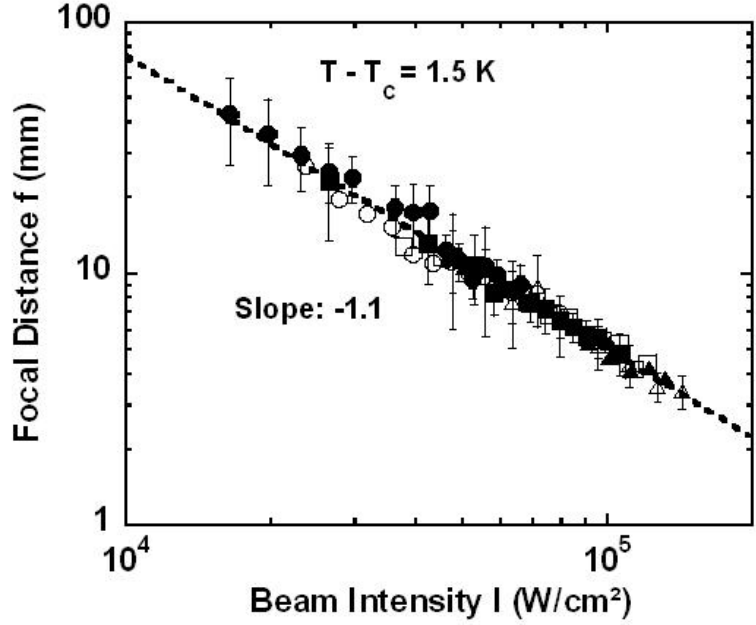

**Figure V.10:** log–log plot of the variation of focal length induced by optical bending versus beam intensity. The dashed curve is the power law fit from Eq. (V.20). The measured slope is $-1.15$ instead of $-1$. The beam waist values are $\omega_0 = 14.6\ \mu m$ (△), $\omega_0 = 18.3\ \mu m$ (▲), $\omega_0 = 21.2\ \mu m$ (□), $\omega_0 = 25.3\ \mu m$ (■), $\omega_0 = 29.3\ \mu m$ (○), $\omega_0 = 32.1\ \mu m$ (●).

## VI – Nonlinear Deformation and "Optohydrodynamic" Instability

In the linear regime in deformation we have assumed that heights were sufficiently weak to be able to linearize the curvature $\kappa$ and to consider optical interaction at normal incidence. However, these approximations obviously fail at larger radiation pressure due to the fact that low surface tension systems, as the one we use are highly deformable. We call this regime, the nonlinear regime in deformation.

### VI.1 – Nonlinear interface deformations

Let us first consider an upward beam propagating from the low to the large refractive index liquid phase. As illustrated in Figure VI.1 the interface shape becomes nonlinear for increasing incident beam powers. For beam power $P \leq 355\ \mu m$ deformations show up a regular bell shape, whereas they stretches out and switches progressively to a surprising tether-like shape of increasing pedestal for $P \geq 475\ \mu m$. These tethers are stable. They do not break into droplets despite their aspect ratio larger than onset for Rayleigh-Plateau instability of liquid columns (see Section VII). Note that nonlinear stretching was already observed in red blood cell deformations [48].

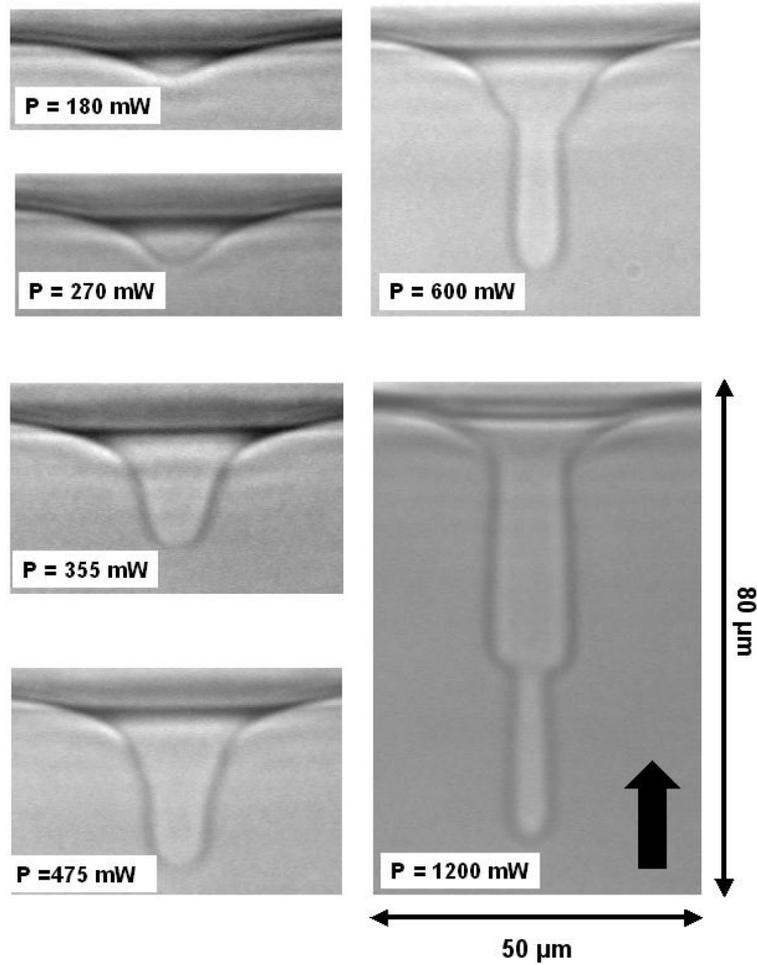

**Figure VI.1:** Nonlinear stationary deformations observed for $T - T_C = 2.5\ K$ and $\omega_0 = 6.3\ \mu m$. The black arrow indicates beam propagation.

To characterize these nonlinear interface deformations, the general equation for the steady interface shape is now required because we cannot assume anymore that $|h'(r)| << 1$ (see Figure VI.1). One has:

$$(\rho_1 - \rho_2)gh(r) - \frac{\sigma}{r}\frac{d}{dr}\left(\frac{rh'(r)}{\sqrt{1+h'(r)^2}}\right) = \frac{n_1 I(r)}{c}\cos^2\theta_1\left(1 + R - \frac{\tan\theta_1}{\tan\theta_2}T\right). \quad \text{(VI.1)}$$

The incidence and transmission angles $\theta_1$ and $\theta_2$ can be related to the shape of the deformation (Figure VI.2) by:

$$\cos\theta_1 = \frac{1}{\sqrt{1+h'(r)^2}}, \qquad \cos\theta_2 = \sqrt{1 - \left(\frac{n_1}{n_2}\right)^2 \frac{h'(r)^2}{1+h'(r)^2}} \quad \text{(VI.2)}$$

As $n_1 \approx n_2$ for near-critical interfaces and $n_1 - n_2 = \Delta n < 0$, Eq. (VI.1) can be further simplified, and we finally obtain:

$$(\rho_1 - \rho_2)gh(r) - \frac{\sigma}{r}\frac{d}{dr}\left(\frac{rh'(r)}{\sqrt{1+h'(r)^2}}\right) = \frac{4\Delta n}{c}\frac{1}{\left(1 + \sqrt{1 - 2\frac{\Delta n}{n_2}h'(r)^2}\right)^2} I(r). \quad \text{(VI.3)}$$

Eq. (V.3) describes the height of the induced deformation whatever the incidence angle is. We assumed a TE wave for the expressions of $R$ and $T$. Due to the cylindrical symmetry of the deformation, a priori such an assumption is debatable because the polarization state is always characterized by a mixing state between TE and TM waves. However, as $\Delta n \to 0$ for near critical interfaces, the difference between TE and TM waves vanishes when approaching the critical point.

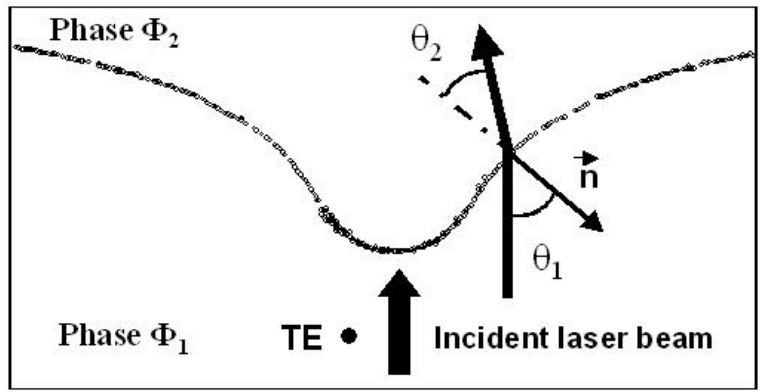

**Figure VI.2:** Definitions of incidence and transmission angles for an upward beam.

To test if Eq. (VI.3) can describe the observed tether-like deformation, we solved it numerically. Comparison between Figure VI.1 and Figure VI.3 shows disagreement since we could not reproduce numerically the observed tether shapes. The proposed reason for such a disagreement is also presented in Figure VI.3, where the beam propagation within the tether can be observed in Inset. While our model supposes that the intensity profile of the exciting beam is always Gaussian, we see that such a hypothesis breaks due to the strong focusing induced by the tip of the deformation and the optical guiding within the tether. The laser wave induces a deformation, which in turn modifies the beam profile.

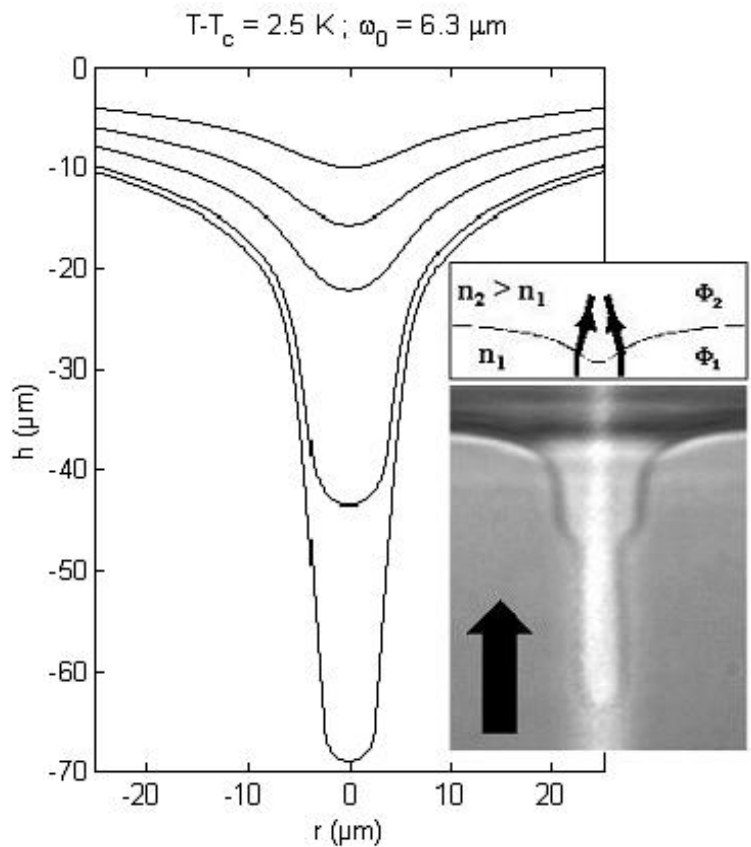

**Figure VI.3:** Numerical simulation of the experiment presented in Figure VI.1 from Eq. (VI.3) for *P = 180, 270, 355, 475, and 600 mW*. Inset illustrates the probable reason for discrepancy between measured and computed interface shapes: the exciting beam within the deformation is no more Gaussian within the tether due to light focusing and guiding.

However, we observe semi-quantitative agreement between the calculated and the measured variations of the tether height versus incident beam power, as illustrated in Figure VI.4.

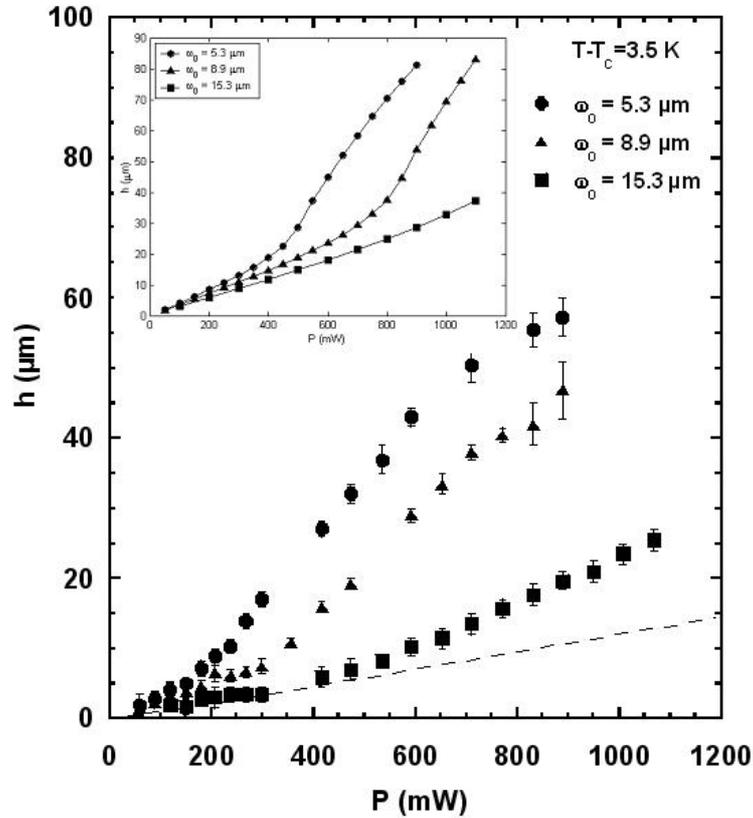

**Figure VI.4:** Variation of the height of the deformation versus incident power in the nonlinear regime for an upward beam (tether formation). Inset: Height predicted by Eq. (VI.3) for the same experimental conditions.

### VI.2 - Up/down symmetry breaking [58]

Let us now consider a downward beam propagating from the large to the low refractive index liquid phase. Comparison of the resulting interface deformation with the upward case is presented in Figure VI.5, while the variations of the height of the deformation versus incident beam power are shown together for both cases in Figure VI.6.

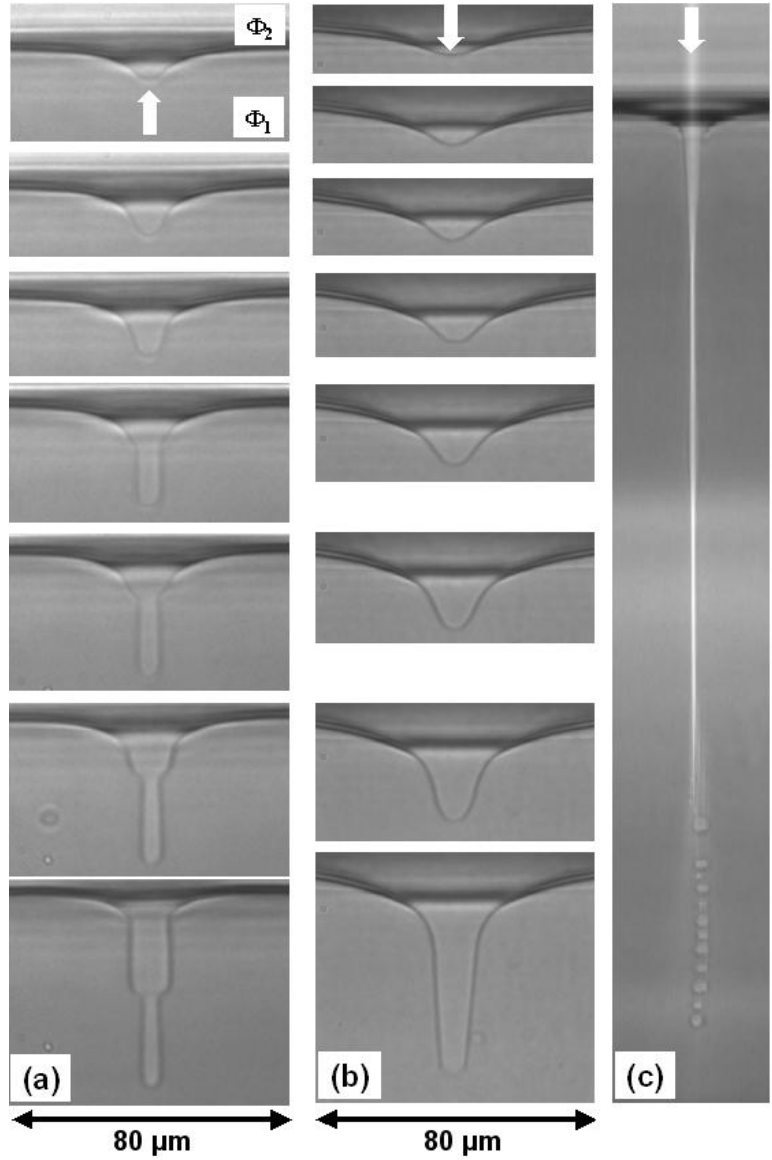

**Figure VI.5:** Interface deformations induced at $(T-T_C) = 3\ K$ with a laser beam waist $\omega_0 = 5.3\ \mu m$. (a) Laser propagating upwards as indicated by the arrow. P increases from top to bottom as $P = 210$, $270$, $300$, $410$, $530$, $590$, and $830\ mW$. (b) Laser propagating downwards as indicated by the arrow. P increases from top to bottom as $P = 190$, $250$, $280$, $310$, $340$, $370$, $400$ and $400\ mW$. The two last pictures are snapshots showing the destabilization of the interface at $P_\uparrow$, leading to the formation of a stationary jet similar to that illustrated in (c) for $(T-T_C) = 6\ K$, $\omega_0 = 3.5\ \mu m$ and $P = 700\ mW$ ($P_\uparrow = 490 mW$). The total height of picture (c) is $1\ mm$.

While for an upward beam we observe the continuous transition from a bell shape to a tether of increasing pedestal, the morphology is totally different when the beam propagates downward from large to low refractive index fluids. In the downward beam case, the deformation height deviates from the linear regime for increasing incident beam powers and diverges at some well-defined beam power threshold $P_\uparrow$. Above this instability onset a beam centered stationary liquid jet forms that emits droplets at its tip.

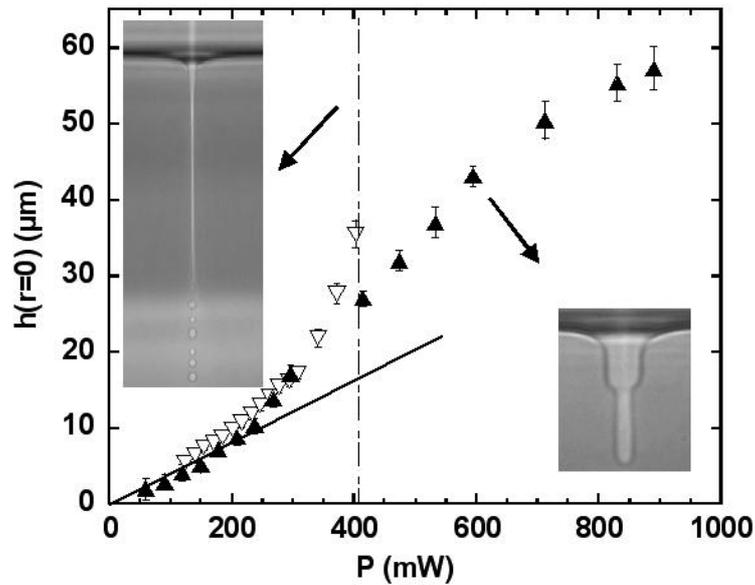

**Figure VI.6:** Evolution of the height of the deformation $h(r=0)$ versus beam power corresponding to the pictures of Figure VI.5a (filled triangles) and Fig VI.5b (empty triangles). The full and broken lines respectively indicate the linear regime in deformation and the threshold value $P_\uparrow$ above which the interface becomes unstable when the laser is propagating downward.

### VI.3 – Interface instability and jet formation [68]

*Interface instability onset*

Interface instability driven by radiation pressure only occurs if the incident fluid has the largest index of refraction. To illustrate its origin, we present in Figure VI.7 pictures showing together the interface and the beam propagation, taken during destabilization for a beam power larger than onset, instead of increasing beam power larger than $P_\uparrow$. At early stage, we can observe the lens effect associated to interface bending. As the deformation grows we further see the beam focalization within the deformation followed by a strong beam trapping and guiding when jet forms. These pictures clearly illustrate the coupling between propagation and deformation. As fluid inside deformation has the largest refractive index, the observed beam focalization could be explained by total internal reflection of light, which in turn would increase light intensity and radiation pressure on the tip deformation, and so on.

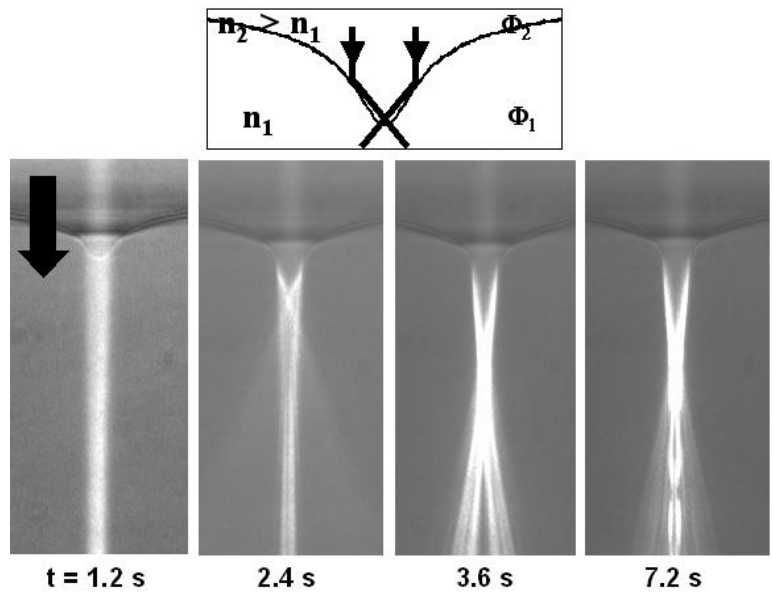

**Figure VI.7:** Proposed mechanism at the origin of interface instability (upper scheme) and jet formation at onset showing optical lensing ($t = 1.2\ s$) followed by total reflection of light at the edge of the deformation ($t = 2.4\ s$ and $t = 3.6\ s$) and the resulting optical self-guiding within the jet ($t = 7.2\ s$). Control parameters are $(T - T_C) = 5\ K$, $\omega_0 = 3.5\ \mu m$ and $P = P_\uparrow = 460\ mW$.

To test the validity of this proposed instability mechanism, we measured the variations of the beam power onset $P_\uparrow$ versus beam waist and temperature. Results are presented in Figures VI.8-9. It appears that $P_\uparrow$ is linear in $\omega_0$ and behaves as $(T - T_C)^\alpha$ with $\alpha$ close to one.

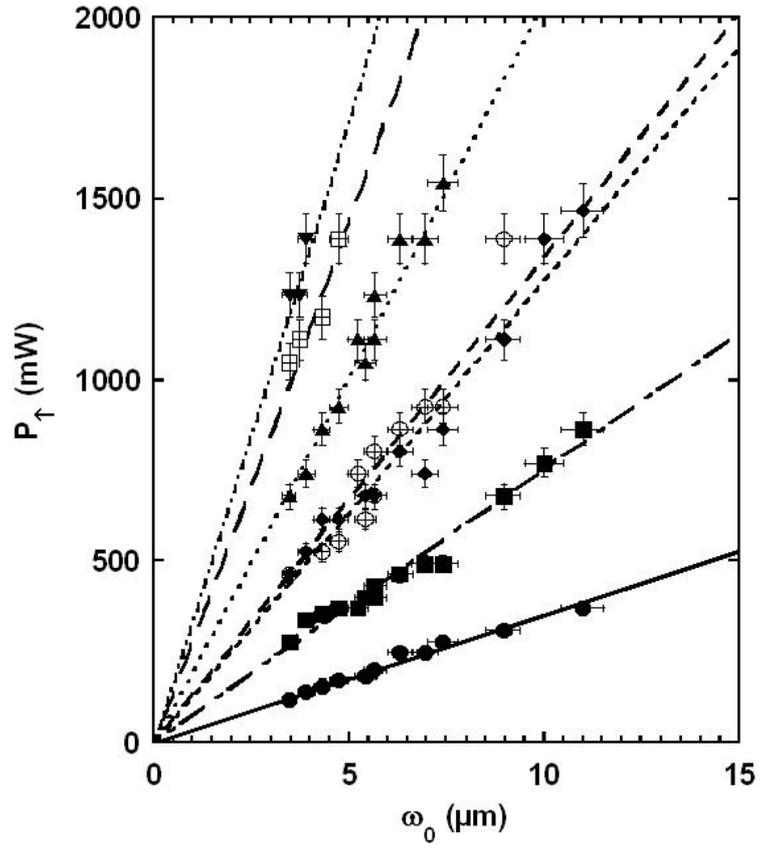

**Figure VI.8:** Variation of the instability onset $P_\uparrow$ versus beam waist $\omega_0$ as a function of $(T-T_C)$. Parameters are: $(T-T_C) = $ *1.5 K* (●), *3 K* (■), *5 K* (◆), *6 K* (○), *8 K* (▲), *12 K* (⊠), and *15 K* (▼). Lines are linear fits.

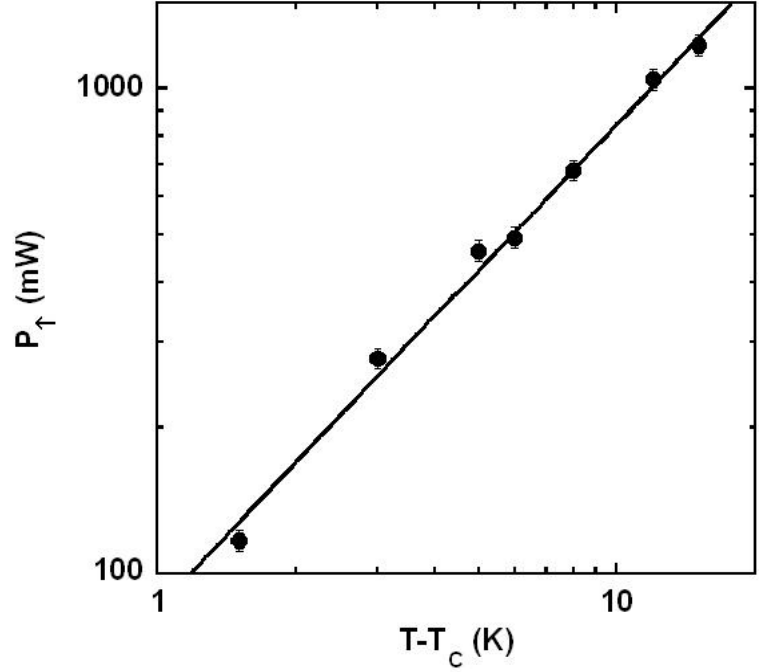

**Figure VI.9:** Variation of the instability onset $P_\uparrow$ versus $(T-T_C)$ for $\omega_0 = 3.5\ \mu m$. The fit gives $P_\uparrow \propto (T-T_C)^{1.01\pm 0.05}$.

To theoretically explain these results we reconsider the general equation that governs the interface shape in stationary conditions (Eq. VI.1), where the incidence and transmission angles, now $\theta_2$ and $\theta_1$, are related to the shape of the deformation (Figure VI.10) by:

$$cos\theta_1 = \sqrt{1-\left(\frac{n_2}{n_1}\right)^2 \frac{h'(r)^2}{1+h'(r)^2}}, \qquad cos\theta_2 = \frac{1}{\sqrt{1+h'(r)^2}}, \qquad (VI.4)$$

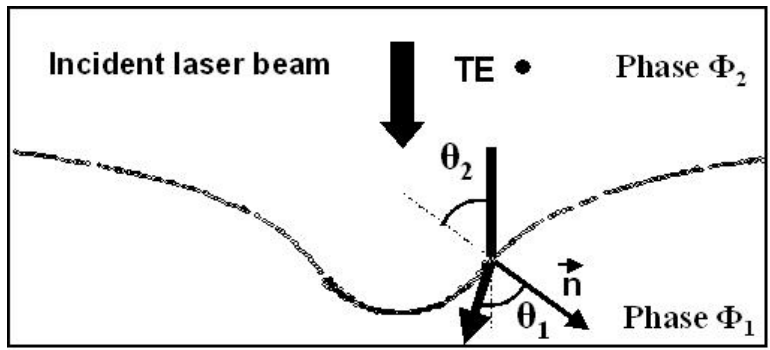

**Figure VI.10:** Definitions of incidence and transmission angles for a downward beam.

As above, we assume $n_1 \approx n_2$ and $n_1 - n_2 = \Delta n < 0$ for our near-critical interfaces. Eq. VI.1 becomes:

$$(\rho_1 - \rho_2)gh(r) - \frac{\sigma}{r}\frac{d}{dr}\left(\frac{rh'(r)}{\sqrt{1+h'(r)^2}}\right) = \frac{4\Delta n}{c}\frac{1}{\left(1+\sqrt{1+2\frac{\Delta n}{n_1}h'(r)^2}\right)^2}I(r). \qquad (VI.5)$$

To further simply equation and get an estimate of the variation of the expected beam power threshold $P_\uparrow$, we develop (Eq. VI.5) at first order in $\Delta n$:

$$(\rho_1 - \rho_2)gh(r) - \frac{\sigma}{r}\frac{d}{dr}\left(\frac{rh'(r)}{\sqrt{1+h'(r)^2}}\right) = \frac{\Delta n}{c}I(r). \qquad (VI.6)$$

As the total reflection condition is $\sin\theta_2 = h'(r)/\sqrt{1+h'(r)^2} \geq n_1/n_2$, integration of Eq. (VI.6) at low Bond number (experiments are realized in this regime where buoyancy $(\rho_1 - \rho_2)gh(r)$ is negligible) leads to:

$$\frac{P|\Delta n|}{2\pi c\sigma}\frac{1}{r}\left[1 - exp\left(-\frac{2r^2}{\omega_0^2}\right)\right] > \frac{n_1}{n_2}. \qquad (VI.7)$$

Total reflection is then obtained for $\sqrt{2}r/\omega_0 = 1.121$, which renders the left hand terms of Eq. (VI.7) maximum. Consequently, the condition for total reflection of light within the deformation defines a beam power threshold as:

$$P \geq \frac{1.121\sqrt{2}\pi}{0.715}\frac{\sigma c}{|\Delta n|}\omega_0, \qquad (VI.8)$$

which is linear in $\omega_0$ and behaves as $(T-T_C)^{2\nu-\beta}$, i.e. with a critical exponent $2\nu - \beta = 0.935$, in very close agreement with the exponent *1* measured for $P_\uparrow$. Eq. (VI.8) also provides a natural scaling to present the disparate data set illustrated in Figure VI.8. This is illustrated in Figure VI.11, where the whole data set is brought onto a single master curve within experimental uncertainties. As expected, $P_\uparrow$ varies linearly in $\sigma c \omega_0 / |\Delta n|$. A linear fit leads to $P_\uparrow(W) = (6.3 \pm 0.3).10^{-6} \left[ \sigma c / |\Delta n| \right] \omega_0 (\mu m)$ while our simple model predicts a slope $6.9 \times 10^{-6}$. This clearly demonstrate that the observed interface instability is triggered by the total reflection of the exciting beam by deformation edges. As illustrated in the following section the resulting beam trapping can be used to build self-adapted optical liquid fibers. Before ending this subsection, we mention that our simplified prediction of the beam power threshold $P_\uparrow$ can be slightly improved if we do not develop radiation pressure at first order in $\Delta n$ [69, 70]; $6.5 \times 10^{-6}$ is theoretically found instead of $6.9 \times 10^{-6}$.

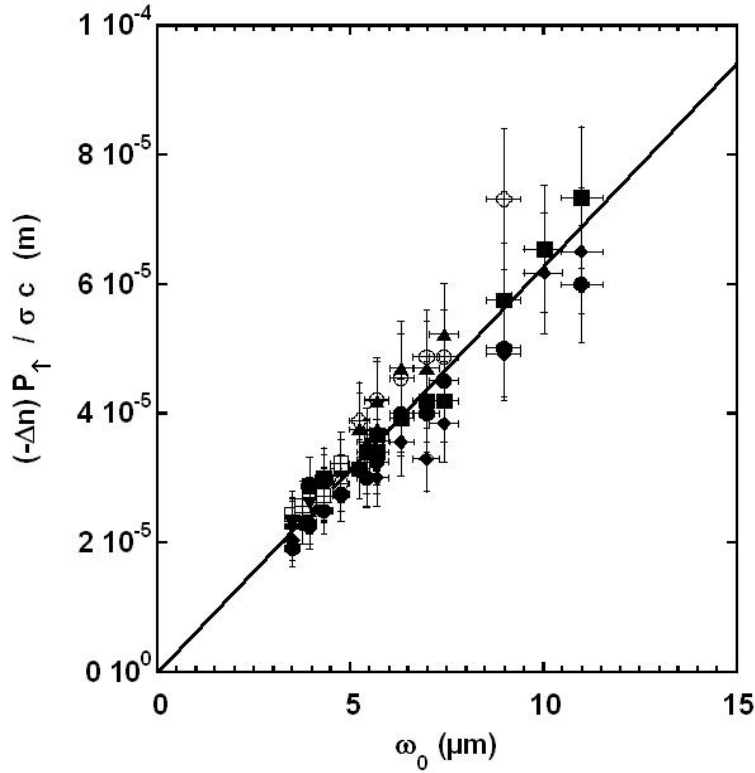

**Figure VI.11:** Rescaling of the set of dispersed data presented in Fig. (VI.8) according to the behavior predicted by Eq. (VI.8). Symbols are the same. The full line is a linear fit.

### Comparison with the Zhang-Chang experiment [36]

Zhang and Chang performed experiments on deformation and rupture of water droplets using high intensity laser pulses. As our model is performed for near-critical interfaces, universality implies that Eq. (VI.8) should be also applicable to the water free surface ($\sigma = 70$ $mN/m$ at room temperature). Moreover, even if the curvature of water droplets can induce further beam focusing on the rear face, Eq. (VI.8) should at least retrieve the right order of magnitude for droplet rupture. We find $P_\uparrow = 29$ $kW$ for a beam waist $\omega_0 = 100$ $\mu m$. Zhang and Chang found that droplet rupture occurs in between *100 mJ* and *200 mJ* laser pulse. However, we have to be careful in directly comparing the associated values to our predicted onset because they used laser pulses instead of continuous waves. It is thus physically more appropriate to perform comparison in term of time dependent intensity rather than beam power [41]. The pulse spatio-temporal profile can be modeled as $I(r,t) = I(r)T(t)$, where:

$$I(r) = \frac{2P}{\pi \omega_0^2} exp\left(-\frac{2r^2}{\omega_0^2}\right), \quad T(t) = \frac{t}{\tau} exp\left(-\frac{t}{\tau}\right)$$

The relaxation time is $\tau = 0.40$ $\mu s$. Taking the effective pulse duration to be $3\tau$, the time average of $T(t)$ is $\langle T \rangle = 0.267$. Consequently, a *100 mJ* pulse with beam waist $\omega_0 = 100$ $\mu m$ gives rise to a plane wave approximated intensity $I_0 = 0.80$ $GW/cm^2$ leading to a time average (over $3\tau$) intensity $I_{pulse} = 0.21$ $GW/cm^2$. Interpretation of $I_{pulse}$ in term of Gaussian profile of intensity $2P/\pi\omega_0^2$, leads to $P_{pulse} = 33$ $kW$, in good agreement with the expected $P_\uparrow = 29$ $kW$ value. However, this value corresponds to a low limit, as Zhang and Chang observed liquid jetting for laser pulses of energy in between *100 mJ* and *200 mJ*. There are also several others reasons that prevent a very accurate comparison. As several data are missing (pulse characterization and accurate measurement of jetting onset, for instance), our calculation is just estimation. Moreover, we did not take into account the curvature of the deformed droplets, and its influence of radiation pressure transfer through transmission and reflection coefficients. Finally, Eq. (VI.8) just describes the onset of the interface instability, while the jet structure is preserved at larger beam power [71]. Nevertheless, even if agreement between predictions and experiments is semi-quantitative, our investigation constitutes the first attempt at comparison on this difficult nonlinear problem.

### *Liquid jet control by optical radiation pressure*

As already illustrated, interface instability at $P_\uparrow$ leads to the formation of a beam centered liquid jet. Despite continuous emission of liquid droplets, the mean jet length is stationary for a given experimental conditions. However, as length of classical jet is controlled by the flow rate, the length of optically driven liquid jets can be continuously tuned by increasing beam power. This is illustrated in Figure VI.12. The first picture corresponds to a beam power *P = 420 mW* smaller than $P_\uparrow$. The fourth following pictures are dynamical views of the development of the jet instability at $P_\uparrow$ = *490 mW*. The jet reaches a stationary length but a continuous emission of droplets from the tip persists. As the distance from the meniscus to the down limit of pictures is *800 μm*, we see that jet length is approximately *300 μm* while typical length for tethers is of the order of a few tens of microns. The interesting point here is that further increase in *P* increases the jet length as illustrated in the following pictures. Consequently, as for the case of linear deformation, one can still have control over the nonlinear regime by simply modifying the characteristics of the exciting beam.

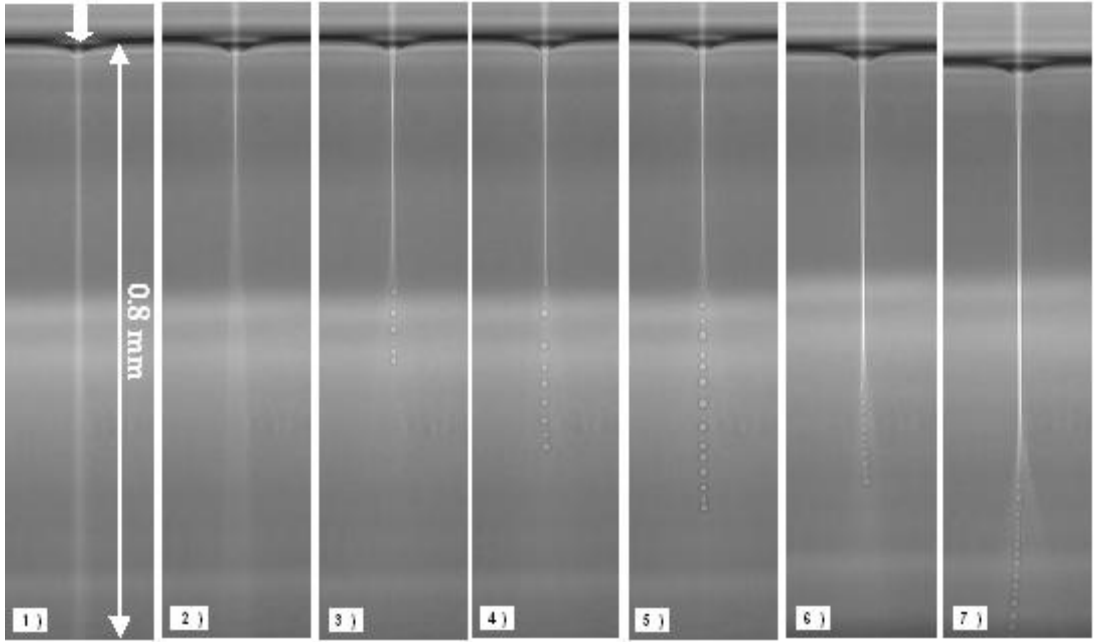

**Figure VI.12:** Liquid jet formation and growth at $(T-T_C) = 6\ K$ induced by a downward beam of beam waist $\omega_0 = 3.5\ \mu m$. 1) Stable deformation at $P = 420\ mW$. 2-5) Dynamics of interface destabilization and jet formation at $P = P_\uparrow = 490\ mW$; note stationary length and continuous droplet flow rate. 6-7) Jet length increase for increasing beam power ($P = 700\ mW$ and $P = 770\ mW$).

**VII – Laser-Sustained Liquid Columns**

While since the beginning we always discussed interfaces deformations induced by radiation pressure within an "optical" context, their nonlinear behaviors are extremely surprising for the point of view of hydrodynamics since liquid fingers of large aspect ratio should not be stable. Indeed, beyond a certain aspect ratio, liquid columns are known to break into droplets due to the Rayleigh-Plateau instability [2]. Noticeable is the fact that laser waves are able to stabilize structures that would be unstable otherwise. This can be easily demonstrated by removing the optical excitation; Rayleigh-Plateau rupture of stationary jets occurs instantaneously. This optical stabilization pushed us to try to stabilize real liquid columns under intense illumination.

*VII.1 - Liquid Bridges and Rayleigh-Plateau Instability*

The standard configuration for liquid bridge analysis is presented in Figure VII.1. A liquid bridge corresponds to a liquid zone of finite volume $V$ surrounded by a second fluid and stabilized between two solid surfaces by surface tension. A parameter that characterizes the bridge is its aspect ratio $\Lambda$ defined as the ratio of its height $L$ over its diameter $2R$: $\Lambda = L/2R$. Such freestanding structure plays a major role in many sciences as different as crystal growth by the floating zone method [72], micropowder adhesion [73], or nanolithography [74]. However, all these processes encounter a fundamental limitation associated to the Rayleigh-Plateau instability [2]. In weightless conditions, a cylindrical liquid column becomes unstable and beaks when its length exceeds its circumference, i.e.:

$$\Lambda = \frac{L}{2R} > \pi. \qquad (\text{VII.1})$$

As Rayleigh-Plateau instability is driven by surface tension, above its onset the liquid column breaks into droplets. Buoyancy even lowers the instability onset given by Eq. (VII.1).

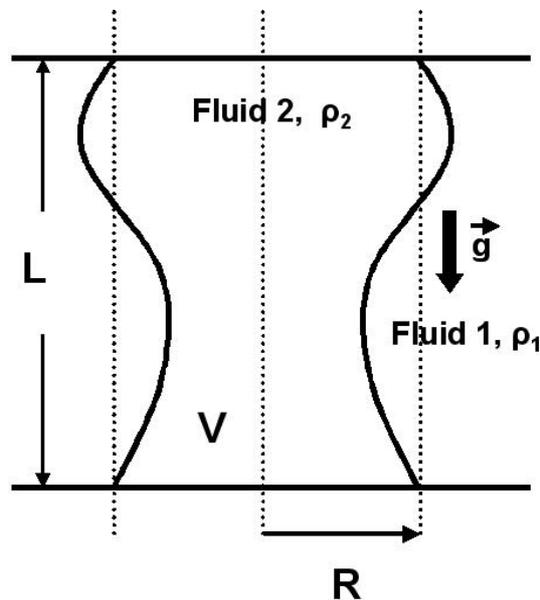

**Figure VII.1:** Classical representation (axial gravity) of a liquid bridge (length $L$, radius $R$ and volume $V$) of fluid *2*, sustained between two rigid boundaries in a fluid *1*. The aspect ratio $\Lambda$ is defined as $\Lambda = L/2R$.

Great deals of efforts have therefore been devoted to bypass this fundamental limitation. Gravity was compensated by magnetic fields [75]. Liquid bridges were stabilized under axial and radial electric fields for both dielectric [76] and conducting [77] liquids. Active control by acoustic radiation pressure was used [78]. The largest value reached was $\Lambda = 5$ [79]. We show in the following subsections that optical radiation pressure helps in largely overshooting this value.

Before studying radiation pressure effects on liquid columns, let us define two important quantities. To analyze liquid bridge stability in presence of buoyancy, i.e. in the most classical situation, one has to compare buoyancy and surface tension effects. As in Section V, this comparison leads to the definition of a classical Bond number:

$$Bo = \frac{\Delta \rho g R}{\sigma / R} = \frac{\Delta \rho g R^2}{\sigma} \qquad (\text{VII.2})$$

As buoyancy compete with surface tension, the Rayleigh-Plateau onset strongly depends on $Bo$ [80]. On the other hand, when an electromagnetic field is used for stabilization, it is also useful to define an electromagnetic Bond number $\chi_E$ as the ratio between electromagnetic and Laplace pressure [76]:

$$\chi_E = \frac{\Delta \varepsilon |E|^2 / 2}{\sigma / R} = \frac{R \Delta \varepsilon |E|^2}{2\sigma} \qquad (\text{VII.3})$$

where $\Delta \varepsilon$ is the contrast in dielectric constant between the two liquids. While $\chi_E$ does not represent any critical value in the problem, it gives nevertheless important insights on stability since $\chi_E > 1$ is required for liquid bridge stabilization above classical Rayleigh-Plateau onset.

### *VII.2 – Optical Stabilization of Liquid Bridges [81]*

The experimental setup used to investigate liquid bridge stabilization under optical radiation pressure is presented in Figure VII.2. A glass capillary is immersed in the optical cell. Its size is chosen so as to be located within the lower fluid phase $\Phi_1$ when the mixture is separated. This capillary plays the role of the two rigid boundaries used to fix the bridge length. As the upper fluid phase $\Phi_2$ wets glass better than $\Phi_1$, we used the unavoidable wetting films that coat the capillary as disconnected reservoirs to form a bridge.

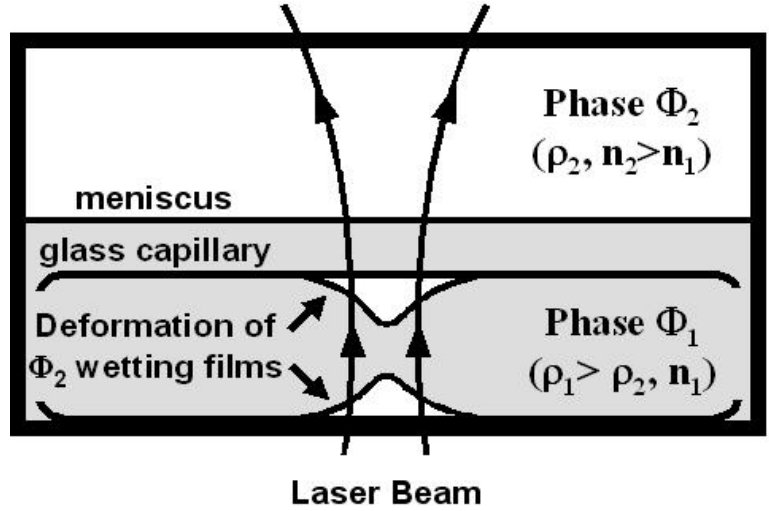

**Figure VII.2:** Schematics of laser-sustained liquid bridge experiment for $T > T_C$. A glass capillary of well-defined height (100 or 200 µm) stays at the bottom of the cell in the less refractive phase $\Phi_1$. Its surface is coated by a wetting film of the most refractive phase $\Phi_2$ that is deformed by the radiation pressure of a focused beam propagating vertically from $\Phi_1$ to $\Phi_2$.

In a second step, we use the general properties of radiation pressure effects on fluid interfaces to deform these wetting films and form a liquid column (Figure VII.3). As interface deformations driven by radiation pressure are always directed towards the less refractive liquid (in our case $\Phi_1$), deformations of both wetting layers stand face-to-face. They are disjoined at low power. The formation of a stable liquid bridge is then ensured by the fact that the lower deformation can become unstable if $P > P_\uparrow$. Two typical examples of liquid bridge formation in a *100 µm* capillary by radiation pressure are presented in Figure VII.3. The first example ($T - T_C = 6\ K$, $\omega_0 = 15.3\ \mu m$) illustrates a situation where the final aspect ratio is close to the classical instability onset $\Lambda = \pi$. The second one ($T - T_C = 15\ K$, $\omega_0 = 3.9\ \mu m$) is chosen so as to also show both the tether formation on the upper deformation and the destabilization of the lower interface. In this case, final aspect ratio is $\Lambda = 12.5$, i.e. well above any already published data. We should notice that interface deformation by a downward beam leads exactly to the same final result, except that now interface instability occurs on the upper deformation. Liquid bridge formation is thus completely independent of the direction of propagation.

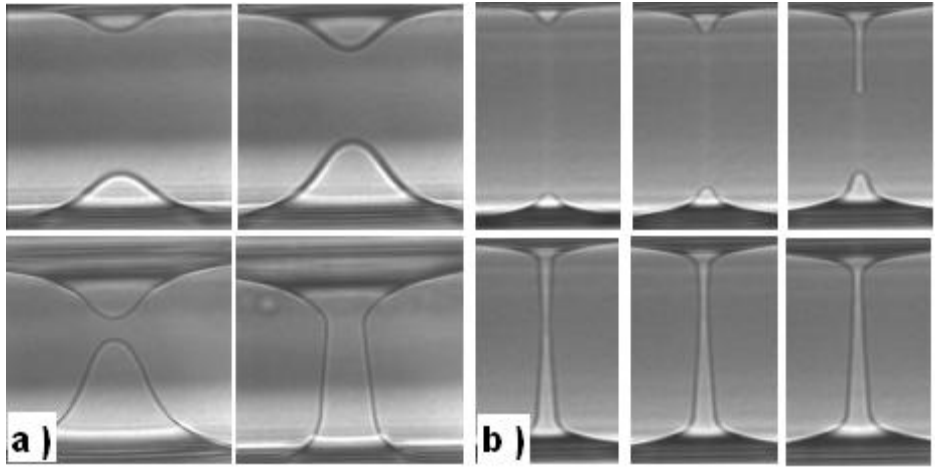

**Figure VII.3:** Liquid bridge formation in a *100 µm*-height glass capillary. (a) $(T-T_C) = 6\ K$ and $\omega_0 = 15.3\ \mu m$. The increasing beam power values from top left to down right are $P = 1500, 2000, 2150\ (P_\uparrow),$ and $2300\ mW$. The resulting liquid bridge has an aspect ratio $\Lambda = 3.6$. (b) $(T-T_C) = 15\ K$ and $\omega_0 = 3.9\ \mu m$. The increasing beam power values from top left to down right are $P = 1150, 1300, 1400\ (P_\uparrow), 1400\ (P_\uparrow), 1600$ and $1900\ mW$. The aspect ratio of the bridge is $\Lambda = 12.5$.

To further characterize these laser-sustained liquid columns, we analyzed the variations of the aspect ratio versus $(T-T_C)$ and $\omega_0$. Results are summarized in Figure VII.4. All the data correspond to stabilized columns of aspect ratio larger than that given from Rayleigh-Plateau criterion. Figure VII.4 shows that $\Lambda$ does not vary with $(T-T_C)$, while it is a strongly decreasing function of $\omega_0$. A power law fit of the full data set leads to $\Lambda = (45 \pm 3) \times \omega_0^{-1.02}$, with $\omega_0$ expressed in $\mu m$. As $\Lambda = L/2R$ and the amplitude factor of the fit is close to the value $L/2 = 50\ \mu m$, one can conclude that the column radius at $P_\uparrow$ is close to $\omega_0$ in confined geometry.

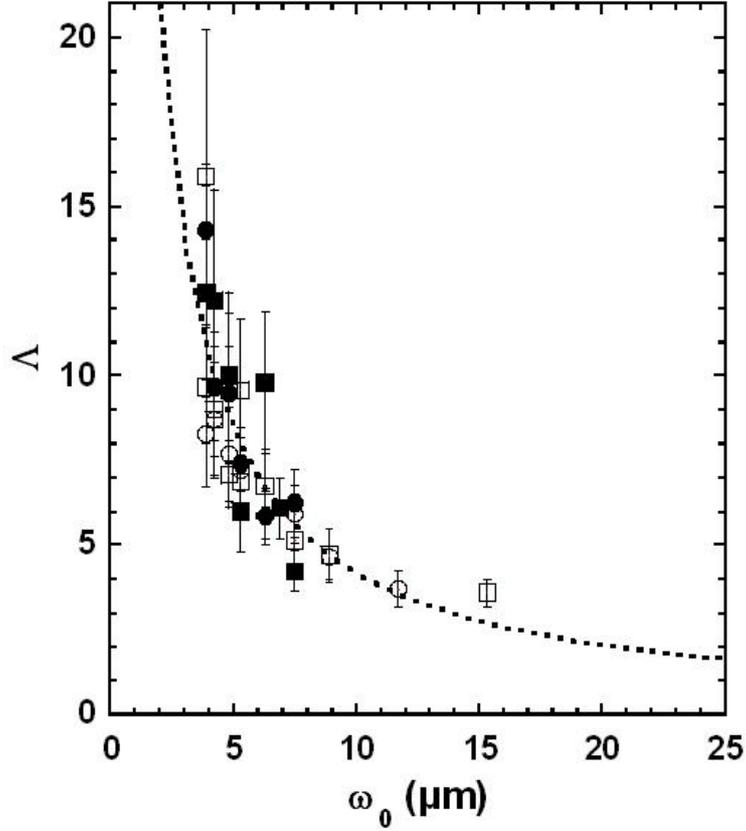

**Figure VII.4:** Liquid bridge stability analysis. Aspect ratio $\Lambda$ versus beam waist $\omega_0$ for liquid bridges stabilized in a 100 μm-height capillary at four different temperatures: $(T-T_C) = 6\ K$ (□), $8\ K$ (○), $10\ K$ (●) and $15\ K$ (■). The full line is a power-law fit of the data leading to $\Lambda \propto \omega_0^{-1.02}$.

To perform an experimental investigation of the liquid bridge stability, we define an electromagnetic Bond number $\chi_{opt}$ that measure the relative importance of optical radiation versus Laplace pressure, as in the electric field case (Eq. VII.3). We get:

$$\chi_{opt} = \frac{2n_1}{c}\left(\frac{n_2-n_1}{n_2+n_1}\right)\frac{2P}{\pi\omega_0^2}\frac{R}{\sigma} \quad \text{(VII.4)}$$

Using Burcham and Saville procedure [76], we first measured the beam power $P_■$ ($\chi_■$) required to form an almost cylindrical bridge. Then, the beam power is lowered until reaching bridge pinch-off at $P_{\blacktriangledown}$ ($\chi_{\blacktriangledown}$). The results, corresponding to the extreme $(T-T_C)$ investigated, are presented in Figure VII.5. Stabilization and pinching of the liquid bridge correspond to well different values of $\chi_{opt}$. $\chi_{\blacksquare}$ and $\chi_{\blacktriangledown}$ are also independent of $(T-T_C)$. This feature can be retrieved by expressing $\chi_{opt}$ as function of $P_{\uparrow}$. Considering the general expression of $P_{\uparrow}$ [68], or its simplified expression close to the critical point (Eq. VI.8) for fluids of close composition, $\chi_{opt}$ can be rewritten as:

$$\chi_{opt} \approx 4\left(\frac{P}{P_{\uparrow}}\right)\left(\frac{R}{\omega_0}\right). \qquad (VII.5)$$

Consequently, as $P_{\blacksquare} \geq P_{\uparrow}$ by definition of $P_{\uparrow}$, and $R \approx \omega_0$ according to Figure VII.4, we recover the fact that $\chi_{\blacksquare}$ should take values close to $4$, as illustrated in Figure VII.5. On the other hand, $\chi_{\blacktriangledown}$ is significantly smaller than $\chi_{\blacksquare}$ due to the fact that the optical guiding within the bridge locally increases field intensity and induces a feedback on the liquid column stability [82]. As already demonstrated by Burcham and Saville in the context of liquid bridge stability under electric field, the dimensionless number $\chi$, in our case $\chi_{opt}$, abstract the main features determining bridge stability, even if the stabilization mechanism remains unknown. A naïve view consists in saying that first there is no radiation pressure coupling acting on a surface of a beam centered cylindrical liquid column. Then, radiation pressure is always directed outward because the index of refraction of the column is by definition larger than that of the surrounding fluid. So, any bridge pinching is automatically balanced by radiation pressure; the more the pinching, the more the contribution of the radiation pressure.

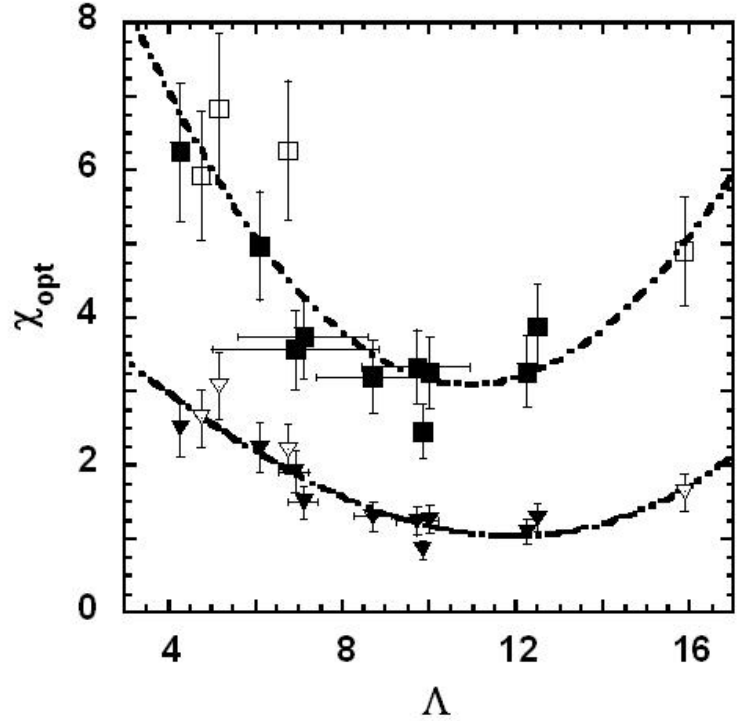

**Figure VII.5**: Stability analysis for the liquid bridges performed at $(T-T_C) = 6\ K$ (empty symbols) and $(T-T_C) = 15\ K$ (filled symbols). The squares represent the values of the dimensionless parameter $\chi_\blacksquare$ for which the bridges are stable and almost cylindrical. The triangles correspond to the values of $\chi_\blacktriangledown$ at which bridge pinching occurs when the beam power is slowly reduced. Dash-dotted lines are guides for the eye.

*VII.3 – Liquid Bridges of Super Large Aspect Ratio Formed by Liquid Jets*

Although the liquid bridges presented in the preceding subsection were stabilized between two rigid boundaries (the edges of a capillary), we can also form liquid columns between the liquid meniscus and the bottom of the experimental cell. Indeed, as illustrated in Figure VI.12, the jet length is controlled by the incident beam power. Thus, as our experimental optical cell is two millimeters high and the meniscus located in the middle due to near criticality, it should be possible to connect the liquid jet to the bottom of the face of the cell and to form a liquid column of super large aspect ratio. As such strategy seems obvious from Figure VI.12 for increasing beam powers, we instead illustrate this liquid column formation dynamically, by choosing a beam power larger than $P_\uparrow$ ($P = 1750\ mW$, $P_\uparrow = 490\ mW$). The dynamics is presented in Figure VII.6. It can be seen that twenty seconds (generally a few tens of seconds) are required to connect the jet to the wetting film that coats the bottom window. As already observed in Figure VI.12 droplets born at the tip of the forming jet ($4\ s \leq t \leq 10\ s$). However, once connection is performed, the liquid column is stable. Its uniformity is surprising as its length is $L \approx 0.8\ mm$ and its radius $R \approx \omega_0 = 3.5\ \mu m$, leading to an aspect ratio $\Lambda = L/2R \approx 110$!

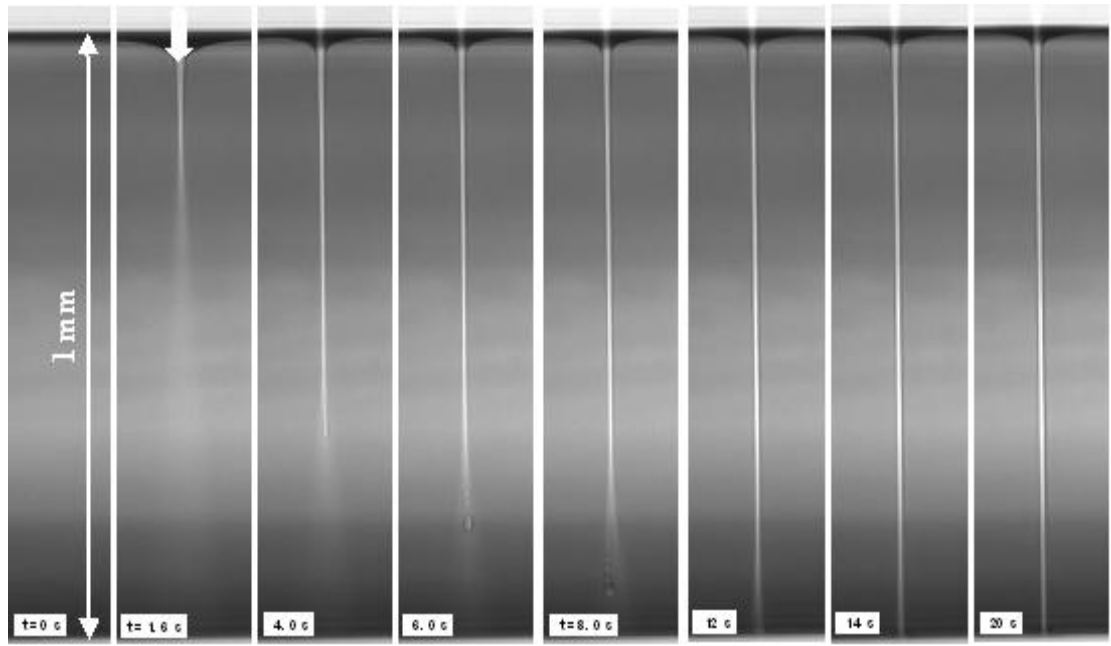

**Figure VII.6:** Dynamics of liquid bridge formation performed at $(T - T_C) = 6\ K$ with a downward beam of beam waist $\omega_0 = 3.5\ \mu m$ and power $P = 1750\ mW$ larger than $P_\uparrow$ ($P_\uparrow = 490\ mW$).

**VIII – Application to Adaptative Optics and Laser Microfluidics**

We already showed in Section V that linear interface deformations could be used for soft lensing with large variations in accessible focal distances. The nonlinear regime in deformation, particularly optically driven liquid jetting, offers an even wider range of application since hydrodynamics starts to couple with light propagation. One major point here is that, contrary to electro-hydrodynamics where micrometric features are difficult to implement, these are natural length scales in "opto-hydrodynamics".

### *VIII.1 - Adaptative Optics*

Beyond soft lensing, the liquid columns stabilized by radiation pressure can be viewed as appealing structures for optical guiding. Indeed, as the index of refraction of the columns is necessarily larger than that of the surrounding fluid, laser beam is automatically captured by total reflection and guided within the liquid medium. This light confinement is dynamically illustrated in Figure VIII.1 over its entire development, i.e. from soft lensing ($t = 1 - 1.5\ s$) to jetting ($t = 2 - 16\ s$) and finally laser guiding ($t = 19 - 22\ s$). The final liquid fiber length is *455 µm*.

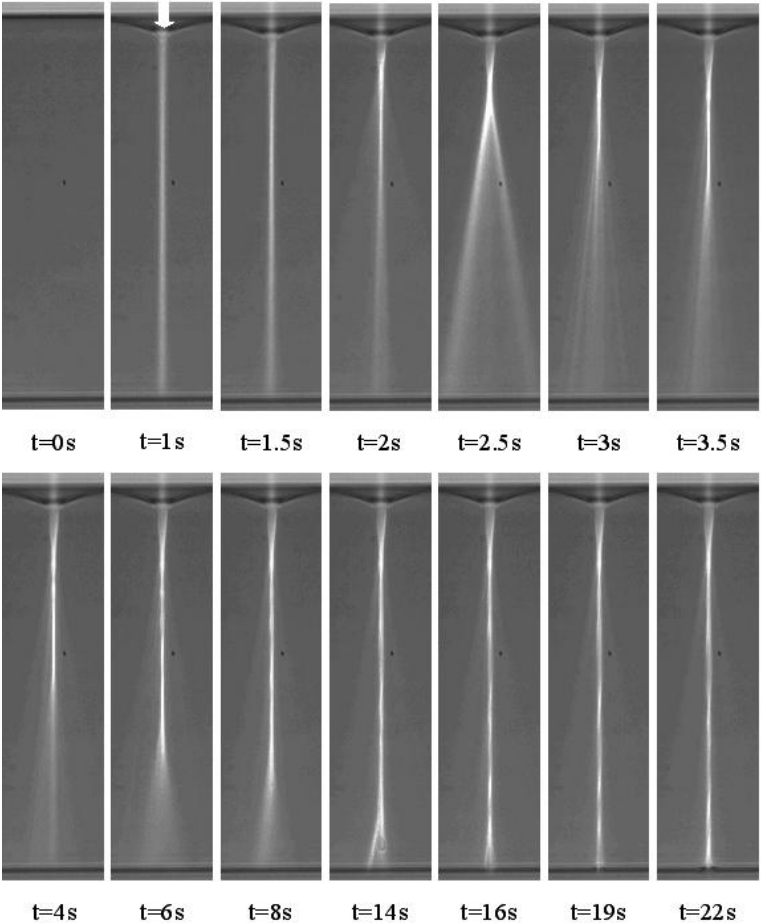

**Figure VIII.1:** Dynamical behavior of soft focusing, laser self-trapping by the induced liquid jet and final optical guiding by the stationary liquid column formed. Parameters are $\omega_0 = 3.47 \ \mu m$, $P = 470 \ mW$, and $(T - T_C) = 4 \ K$. The liquid fiber length is $455 \ \mu m$.

This laser self-guiding is also efficient in smaller liquid bridges as illustrated in Figure VIII.2a, where a liquid column of aspect ratio $\Lambda = 14$ is stabilized with a glass capillary of height $200 \ \mu m$.

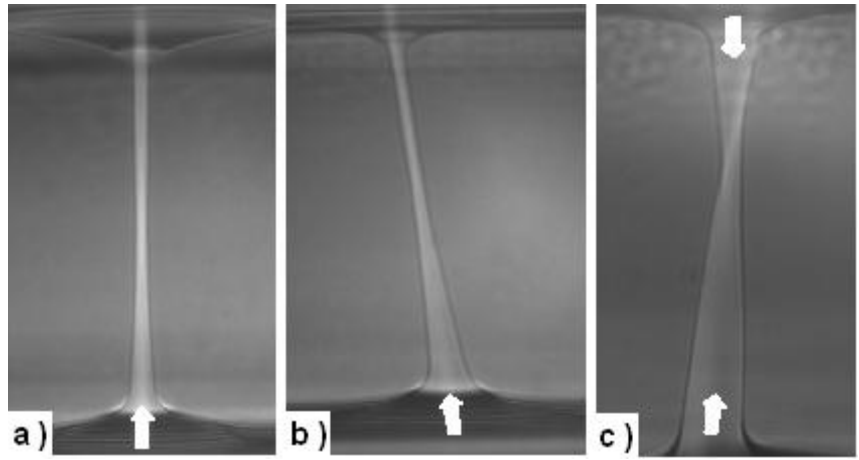

**Figure VIII.2:** a) Liquid bridge with aspect ratio $\Lambda = 14$ induced and stabilized in a $200 \ \mu m$ glass capillary by a laser beam of beam waist $\omega_0 = 3.2 \ \mu m$ and power $P = 1400 \ mW$. The temperature gap is $(T - T_C) = 6 \ K$. The brightness of the column evidences laser self-guiding. b) Same experimental conditions as in a) but with a tilted exciting beam. c) Liquid elbow created in a two-laser configuration (upward/downward) at $(T - T_C) = 6 \ K$. Both wetting films are destabilized, intercept and form a wedge. The directions of propagation of each beam ($\omega_0 = 3.5 \ \mu m$, $P = 800 \ mW$) are indicated by the arrows.

One can therefore put forward the concept of liquid step-index optical fibre, which provides a very new approach towards self-induced waveguiding. Indeed, contrarily to optical adaptative waveguides written by photopolymerisation [83] or laser damage in glasses [84], liquid fibres are non-permanent. They are thus completely reconfigurable. The laser creates its own channel that is automatically optimised to its waist and power as illustrated in Figure VIII.3, where adaptation to incident beam power is clearly observed.

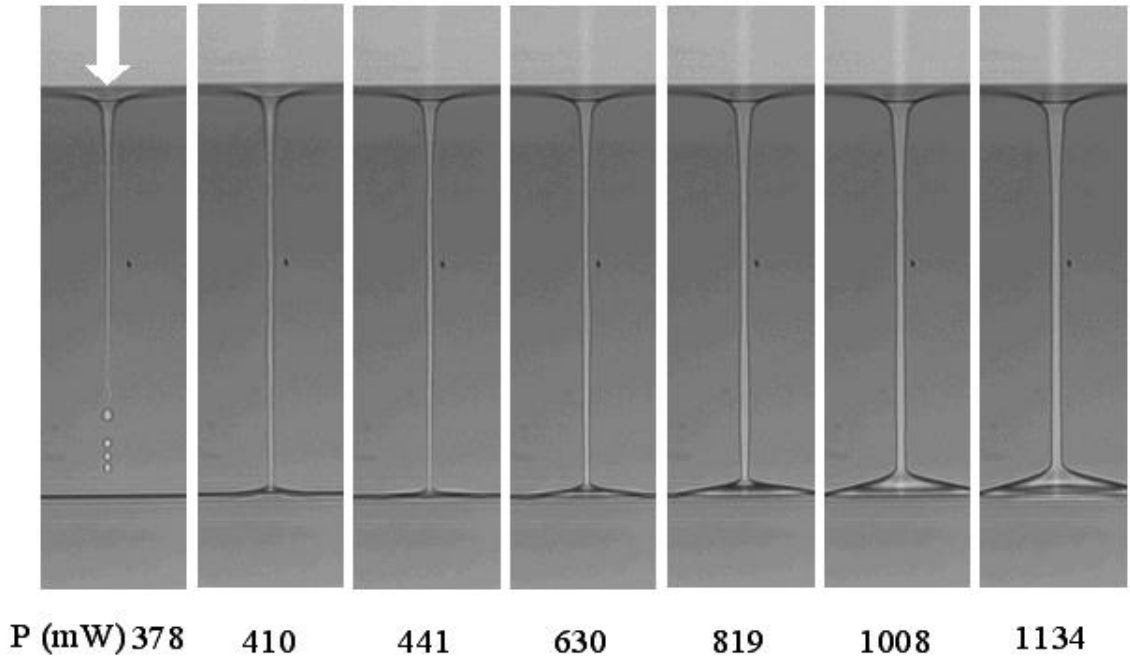

**Figure VIII.3:** Adaptation of the liquid fiber diameter to the input power. Parameters are $\omega_0 = 3.47~\mu m$ and $(T - T_C) = 4~K$. The liquid fiber length is $334~\mu m$. As demonstrated by the emission of droplets from liquid jets, note that hydrodynamic flow still persists within the liquid columns as illustrated by the fluid accumulation at their feet.

These tunable optical fibres can furthermore be oriented in any direction by tilting the exciting beam. Figure VIII.2b gives an example of such an inclined liquid bridge stabilized by a laser beam propagating obliquely. Even more surprising are artificial structures such as the stable liquid elbow created in a two-laser beams configuration and presented in Figure VIII.2c. Optically induced liquid columns are thus particularly efficient to control beam propagation or to optimise light coupling devices because self-adaptation considerably reduces the sensitivity to precise mechanical alignments of the optical components used.

*VIII.2 – Laser Microfluidics*

Aside optical guiding, the liquid jets and columns stabilized by radiation pressure can be viewed as three dimensional microfluidic devices with "soft" wall because liquid is transported from the upper liquid phase when $P \geq P_\uparrow$, i.e. above interface instability onset. Two types of micro-flow can be generated: droplet flow and pipe flow.

As illustrated dynamically in Figure VIII.4, droplets are continuously generated at the tip of a stationary jet. Moreover, since the index of refraction of the droplet is larger than that of the surrounding fluid, the beam automatically traps them. This brings directionality in droplet emission and transport that can be actuated by tilting the beam as illustrated in Figure VIII.2b. Finally, at fixed $(T - T_C)$ and $\omega_0$, the droplet flow rate can be controlled by the incident beam power. Consequently, contrary to other methods [85], this optical approach of microfluidics provides droplets that are directly produced within the chosen micrometric size in a contact less way without further processing. Also, as droplets are produced in three dimensions from a channel that can be controlled in size by the beam, no particular microfluidic device is required to manage the droplet flow.

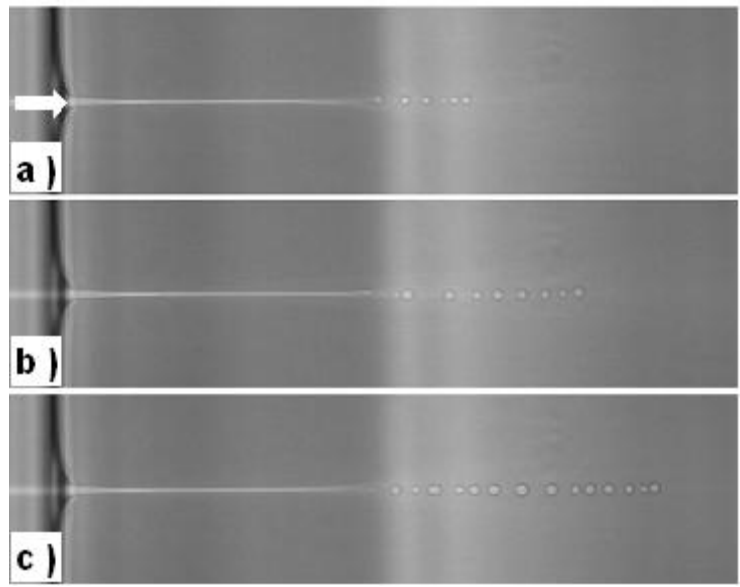

**Figure VIII.4:** Continuous droplet emission and trapping from a jet of stationary length. Note the regularity in flow rate and the relative monodispersity of the induced droplets. The time delay between two pictures is *1 s* and the jet length is *330 μm*. Control parameters are $(T - T_C) = 6\ K$, $\omega_0 = 3.5\ \mu m$, and $P = P_\uparrow = 490\ mW$.

On the other hand, liquid flow within the jet persists when it reaches the bottom face of the cell and forms a liquid bridge. This is illustrated in Figure VIII.3, were accumulation of liquid on the bottom window of the cell is clearly observed. This means that laser-sustained liquid bridges behave as micro-pipes that can be used to transfer fluid from one reservoir to another. One even more surprising fact is that the micro-pipe section can be actuated continuously by varying the incident beam power (see Figure VIII.3).

Finally, let us note that liquid columns are unstable in classical conditions. As column breaking generally leads to at least bimodal droplet distributions, often called droplets and satellites [2] (see the left picture of Figure VIII.5, the important drawback for microfluidic applications is control over micro-jet fission to form a regular assembly of micro-droplets [86, 87]. On the other hand, for applications motivated by fluid transport alone annular flows are much more efficient than droplet flows, even if the first situation is difficult to implement due to the Rayleigh-Plateau instability. Since radiation pressure of laser waves is able to prevent column breaking, it becomes possible to reverse the microfluidic approach in order to transform a bubbly flow into an annular flow. This process is dynamically illustrated in Figure VIII.5. Starting from a linear assembly of well-separated droplets, the radiation pressure deforms droplets and forces them to coalesce transforming the initial droplet assembly into a liquid column.

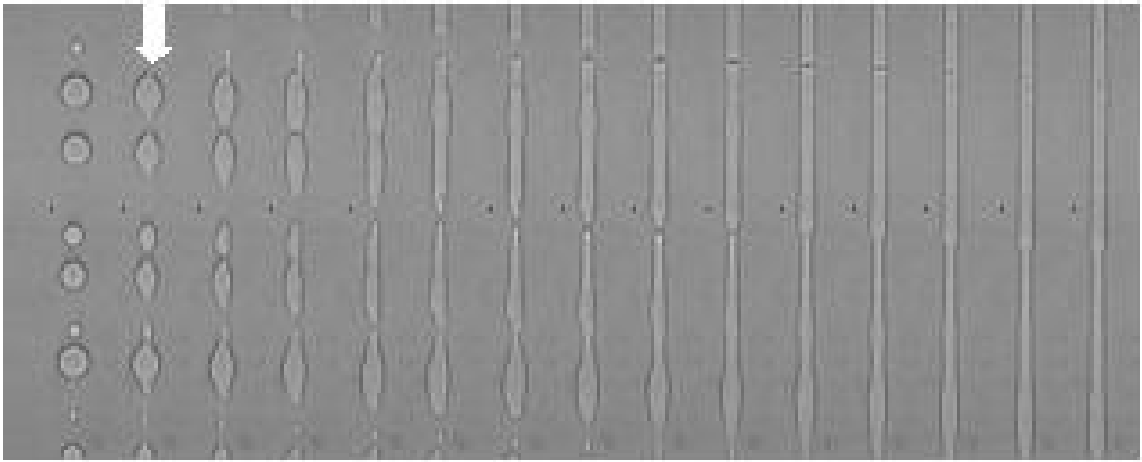

**Figure VIII.5:** Transition from droplet to annular flow resulting from droplet deformation and coalescences driven by the optical radiation pressure. The acquisition frequency is $f = 20\ Hz$. Control parameters are $(T - T_C) = 5\ K$, $\omega_0 = 3.47\ \mu m$, and $P = 630\ mW > P_\uparrow$.

## IX – Conclusion and Prospects

The main purpose of this review was to theoretically and experimentally explore fluid interface deformation driven by the optical radiation pressure of a continuous laser wave. Using near-critical liquid-liquid interfaces, to strongly reduce surface tension and enhance radiation pressure, we validated a universal description of the process.

We also investigated nonlinear regimes in deformation. Asides surprising stable tether shapes, we have presented a new electromagnetic instability mechanism of fluid interfaces driven by continuous laser wave. The very good agreement observed between measurements and expectations demonstrates the universality of this nonlinear process. The resulting jet is also expected to occur quite universally and its regularity, as well as that of the produced micro-droplets, should be promising in microfluidics and electromagnetic spraying techniques.

We finally extended this investigation on nonlinear behaviors to the formation of stable liquid columns. While the results presented here were obtained with a single laser beam, the method could easily be extended to a parallel approach by forming liquid-bridge patterns or adaptative liquid gratings by tailoring the intensity distribution with interfering pump beams.

Even if we try to give an extended view of laser-induced fluid interface deformation, most of its theoretical description, particularly in the nonlinear regime, is still missing. For example, we did not explain the observed tether shape of the deformation. An investigation of liquid jet properties is also missing. The main reason is linked to the complex nonlinear coupling between a mobile and deformable interface and a laser beam as the interface adapts its position to optical excitation. Also, we did not discuss dynamical behaviors, even in the linear regime in deformation. Even if we do have some results on this important aspect, particularly for applications, they are too incomplete to be developed within a general scheme.

Nevertheless, we hope that our exhaustive presentation of the "static" manifestations of a liquid interface under the radiation pressure of a laser wave will promote an optical approach to interface actuation and, to build microfluidic devices with optical forces [88] or, conversely, to anticipate new optical micro-systems based on microfluidics [89] for flow guiding and light coupling applications.

**Acknowledgments**

We are grateful to J. Plantard, M. Winckert, W. Benharbone, C. Lalaude and T. Douar for technical assistance. This work was partly supported by the CNRS and the Conseil Régional d'Aquitaine.